\documentclass{aa}  

\usepackage{graphicx}
\usepackage{txfonts}
\usepackage[normalem]{ulem}

\newcommand{\einstein}{{\it Einstein}}
\newcommand{\rosat}{{\it ROSAT}}
\newcommand{\chandra}{{\it Chandra}}
\newcommand{\xmm}{{\it XMM-Newton}}
\newcommand{\suzaku}{{\it Suzaku}}
\newcommand{\akari}{{\it AKARI}}

\newcommand{\hubble}{{\it Hubble}}
\newcommand{\spitzer}{{\it Spitzer}}
\newcommand{\fermi}{{\it Fermi}}

\newcommand{\tbvarabs}{{\tt tbvarabs}}
\newcommand{\apec}{{\tt apec}}
\newcommand{\vapec}{{\tt vapec}}
\newcommand{\vnei}{{\tt vnei}}
\newcommand{\pow}{{\tt powerlaw}}
\newcommand{\kt}{{$kT$}}
\newcommand{\net}{{$n_e t$}}
\newcommand{\cms}{cm$^{-3}$ s}
\newcommand{\nh}{{$N_{\mathrm{H}}$}}
\newcommand{\cm}{cm$^{-2}$}

\newcommand{\hi}{$\ion{H}{i}$}
\newcommand{\hii}{$\ion{H}{ii}$}

\newcommand{\nbr}[1]{\left( #1 \right)} % normal bracket
\newcommand{\ff}[2]{\nbr{\frac{#1}{#2}}}
\newcommand{\Gf}{\Gamma_\mathrm{f}}

\begin{document}

   \title{eROSITA studies of the Carina Nebula}

   \titlerunning{}

   \subtitle{}

   \author{Manami Sasaki
          \inst{1}
          \and
          Jan Robrade
          \inst{2}
          \and
          Martin G. H. Krause
          \inst{3}
          \and
          Jonathan R. Knies
          \inst{1}
          \and
          Kisetsu Tsuge
          \inst{1}
          \and
          Gerd P\"uhlhofer
          \inst{4}
          \and
          Andrew Strong
          \inst{5}
          }

   \institute{
Dr.\ Karl Remeis Observatory, Erlangen Centre for Astroparticle Physics, Friedrich-Alexander-Universit\"{a}t Erlangen-N\"{u}rnberg, Sternwartstra{\ss}e 7, 96049 Bamberg, Germany \email{manami.sasaki@fau.de}
\and
Hamburger Sternwarte, Universit{\"a}t Hamburg, Gojenbergsweg 112, 21029 Hamburg, Germany
\and
Centre for Astrophysics Research, School of Physics, Astronomy and Mathematics, University of Hertfordshire, 
College Lane, Hatfield, Hertfordshire, AL10 9AB, UK
\and
Institut f{\"u}r Astronomie und Astrophysik, Universit{\"a}t T{\"u}bingen, Sand 1, 72076 T{\"u}bingen, Germany
\and
Max-Planck-Institut f\"{u}r extraterrestrische Physik, Gie{\ss}enbachstra{\ss}e 1, 85748 Garching, Germany
%\email{aws@mpe.mpg.de}
             }

   \date{Received June 11, 2023; accepted December 03, 2023}

% \abstract{}{}{}{}{} 
% 5 {} token are mandatory
 
  \abstract
   {
During the first four all-sky surveys eRASS:4 carried out from December 2019 to 2021, the extended Roentgen Survey with an Imaging Telescope Array (eROSITA) on board Spektrum-Roentgen-Gamma (Spektr-RG, SRG) observed the Galactic \hii\ region Carina nebula.}   
   {We analysed the eRASS:4 data to study the distribution and the spectral properties of the hot interstellar plasma and the bright stellar sources in the Carina nebula.}
   {Spectral extraction regions of the diffuse emission were defined based on X-ray spectral morphology and multi-wavelength data. The spectra were fit with a combination of thermal and non-thermal emission models. X-ray bright point sources in the Carina nebula are the colliding wind binary $\eta$ Car, several O stars, and Wolf-Rayet (WR) stars. We extracted the spectrum of the brightest stellar sources, which can be well fit with a multi-component thermal plasma model.}
   {The spectra of the diffuse emission in the brighter parts of the Carina nebula is well reproduced by two thermal models, a lower-temperature component ($\sim$0.2 keV) and a higher-temperature component (0.6 -- 0.8 keV). An additional non-thermal component dominates the emission above $\sim$1 keV in the central region around $\eta$ Car and the other massive stars. Significant orbital variation of the X-ray flux was measured for $\eta$~Car, WR~22 and WR~25. $\eta$ Car requires an additional time-variable thermal component in the spectral model, which is associated to the wind-wind-collision zone. }
   {Properties like temperature, pressure, and luminosity of the X-ray emitting plasma in the Carina nebula derived from the eROSITA data are consistent with theoretical calculations of emission from superbubbles. It confirms that the X-ray emission is caused by the hot plasma inside the Carina nebula which has been shocked-heated by the stellar winds of the massive stars, in particular, of $\eta$ Car.}


   \maketitle
%
%-------------------------------------------------------------------
\section{Introduction}

The Carina nebula complex (NGC 3372, hereafter CNC) is the 
brightest emission nebula in the southern Galactic plane 
at a distance of 2.3$\pm$0.1 kpc
\citep{2006MNRAS.367..763S,2021ApJ...914...18S,2022A&A...660A..11G}.
It harbours hundreds of protostars and 
several young stellar clusters (Trumpler 14, 15, and 16; 
Collinder 228 and 232; and Bochum 10 and 11) and is a
giant \hii\ region, which has been created by the radiation and winds from more 
than 60 OB stars, three Wolf-Rayet (WR) stars, and the massive binary 
$\eta$ Carinae ($\eta$ Car hereafter) 
in Trumpler 16 
\citep{2008hsf2.book..138S}.
In addition, at about $1^\circ$ northwest of the CNC, an \hii\ region around 
the stellar cluster NGC 3324 called the
Gum 31 region is located, with a  similar distance as to the CNC 
\citep{2005A&A...438.1163K}.
Thanks to its proximity, the CNC is the ideal place to study star formation and 
massive star feedback in the Milky Way.

Inside the CNC, there are dark shells observed in the optical, in which stars are being formed \citep{2000ApJ...532L.145S,2010MNRAS.406..952S}.
Images taken with the \hubble\ and \spitzer\ Space Telescopes have revealed 
large pillars of dust in the nebula around $\eta$ Car.
A large amount of dust has also been found in observations in the sub-mm-range 
\citep{2011A&A...525A..92P} and the far-infrared \citep[FIR,][]{2012A&A...541A.132P}.
Observations of molecular line emission have revealed some dynamics in the 
dust shells with velocities that are different from that of  
the Carina arm, in which they are located
\citep{1984PASA....5..552W,1995A&A...299..583C,1998PASA...15..202B,2005ApJ...634..476Y}. 

In the CNC, there is also molecular gas which may develop into 
new star-forming regions in the future \citep{2008hsf2.book..138S}.
Three regions with different star formation properties have been
detected in the CO data: 
Near $\eta$ Car, there are small globules, which are most likely surviving 
cores in the giant molecular cloud complex. 
Around Trumpler 14, there is the Northern Cloud, which shows little indication for star formation. 
In the south, there is the Southern Cloud, which has been affected
by the massive stars in Trumpler 16 and in which a second generation of stars 
is forming.

Active star formation is in particular observed in the dust pillars. In optical
and infrared (IR) observations, more than 900 young stellar objects (YSOs) 
have been detected in the dust pillars \citep{2010MNRAS.406..952S}
including protostars with ages of only 10$^5$ yr. The star formation in these dust 
pillars was most likely triggered by winds from the massive stars.
In the entire CNC, more than 60,000 young stars have been detected
\citep{2011A&A...530A..34P}.
With such a high rate of star formation, the CNC contributes significantly to the current star formation in the Milky
Way \citep{2011ApJS..194...14P,2011A&A...530A..34P}.

Soft X-ray emission has also been observed from the CNC with the
\einstein\ Observatory \citep{1982ApJ...256..530S}
and \rosat\ \citep{1995ApJ...445L.121C}.
Observations with \xmm\ and \suzaku\ have shown a thermal spectrum with
enhanced element abundances particularly of iron 
\citep{2007PASJ...59S.151H,2009PASJ...61S.123E}.
With the \chandra\ X-ray Observatory, diffuse X-ray emission was confirmed
and analysed in a 1 Ms survey of the CNC 
\citep[Chandra Carina Complex Project, CCCP,][]{2011ApJS..194...15T,2011ApJS..194....1T}.
Spectral analysis of the diffuse emission has shown that the emission is
from thermal plasma, with indications for charge exchange, most likely
between the hot plasma inside the nebula and the cold pillars, ridges, and 
clumps. The diffuse X-ray emission is most likely caused by
the stellar winds of massive stars \citep{1975ApJ...200L.107C}, 
however, observations have also shown, that the 
X-ray emitting plasma is not concentrated around the massive stars but 
is distributed in more broad areas.
There is a bright `hook' in X-rays, which does not coincide with any prominent emission at longer wavelengths. In the east, the emission of the `X-ray hook' is aligned with the eastern arm of the striking V-shaped dust lane seen in the optical.
The diffuse X-ray emission is brightest south of the western arm of the V-shaped dust lane.
Some anti-correlation is seen between the diffuse X-ray emission and the 
8 $\mu$m IR emission, 
suggesting that the X-rays shine through holes in the distribution of 
heated dust. Alternatively, the bulk of the X-ray plasma is too hot and too tenuous to be seen, such that it only appears where that hot plasma pushes through the holes and mixes with  gas from the hole rim. The mixing of both reduces the temperature of the hot gas and increases its emission measure, and the gas appears in soft X-rays. This process is predicted from 3D hydrodynamic simulations to occur frequently whenever massive-star wind bubbles merge.

In this paper we report on the studies of the diffuse emission of the CNC  using the extended ROentgen survey with an Imaging Telescope Array  \citep[eROSITA;][]{2012arXiv1209.3114M,2021A&A...647A...1P} 
on board the Spektrum-Roentgen-Gamma (Spektr-RG, SRG) spacecraft.
eROSITA has performed all-sky surveys in the soft to medium X-ray energy
band of 0.2 -- 10 keV, which is in particular suited for the study of
the hot phase of the interstellar medium (ISM). The all-sky surveys allow us to study the emission from the entire nebula with a uniform exposure and resolution. In addition, as the emission in the surroundings is observed as well, we are able to constrain the local X-ray background well. We present the study of the morphology and the spectra of the diffuse emission in different regions. We compare the physical parameters of the ISM derived from the eROSITA data with theoretical calculations.
We have also analysed the emission of the X-ray brightest massive stars in the CNC and study the time variability.

%--------------------------------------------------------------------
\section{Data}

\subsection{eROSITA}

The Carina nebula complex was observed in the all-sky surveys 
of eROSITA on SRG, which was launched on July 13, 2019 from the 
Baikonur cosmodrome.
There are seven telescope modules (TMs) on eROSITA, 
which are each equipped with a CCD detector. 
After the launch, it became obvious that two TMs suffer light leak 
(TM5 and 7) and therefore have a higher background. 
The effect is not well understood yet, therefore, data taken with TM5 and 7 
are not included in the analysis presented here.
eROSITA has a large field of view (FOV) with a diameter of $\sim$1 degree and 
a high sensitivity in the energy range up to 10 keV, especially below 2.3 keV.
It is thus the ideal telescope to study the distribution and the 
spectral properties of the large extended X-ray emission from the ISM.
eROSITA is planned to perform a total of eight surveys of the entire sky 
(eROSITA all-sky survey, eRASS1-8, the total survey called eRASS:8), out of 
which four surveys have been completed now (eRASS:4). We present the analysis 
of the eRASS:4 data of the CNC.

For data processing and analysis, we have used the 
eROSITA Science Analysis Software System 
\citep[\texttt{eSASS},][]{2022A&A...661A...1B}.
We have used the energy calibrated event files from the eSASS pre-processing 
pipeline c020 and eSASS users version 211214.

The Carina nebula complex has an extent of about 2$^\circ \times$ 3$^\circ$. 
In order to fully cover the CNC and also to have data for estimating the
local X-ray background, we have used eRASS:4 data from a region with a size of 
$\sim10^\circ \times 5 ^\circ$ centered on the CNC.

\subsection{HI}
Archival data of H{\sc i} 4$\pi$ (HI4PI) all-sky H{\sc i} survey \citep{2016A&A...594A.116H} were used in this work to compare the distribution of the hot interstellar plasma emitting X-rays with that of the cold diffuse gas. It is likely that the massive stars in the CNC have produced a cavity in the distribution of the cold atomic gas, which is filled with photo- and shock-ionised gas seen in the optical to X-rays. The HI4PI survey combines the Effelsberg-Bonn H{\sc i} Survey \citep[EBHI;][]{2011AN....332..637K,2016A&A...585A..41W} and Galactic All-Sky Survey \citep[GASS;][]{2009ApJS..181..398M,2010A&A...521A..17K,2015A&A...578A..78K} obtained with the Parkes telescope. The angular resolution is 16.$\arcmin$2, corresponding to $\sim$ 11 pc at the distance of the Carina nebula \citep[2.3 kpc;][]{2006MNRAS.367..763S}. The brightness temperature noise level is $\sim$43 mK at a velocity resolution of 1.49 km s$^{-1}$.

%-----------------------------------------------------------------
\section{Diffuse emission}

\subsection{Images}\label{ana:ima}

\begin{figure*}
\begin{center}
\includegraphics[width=\textwidth]{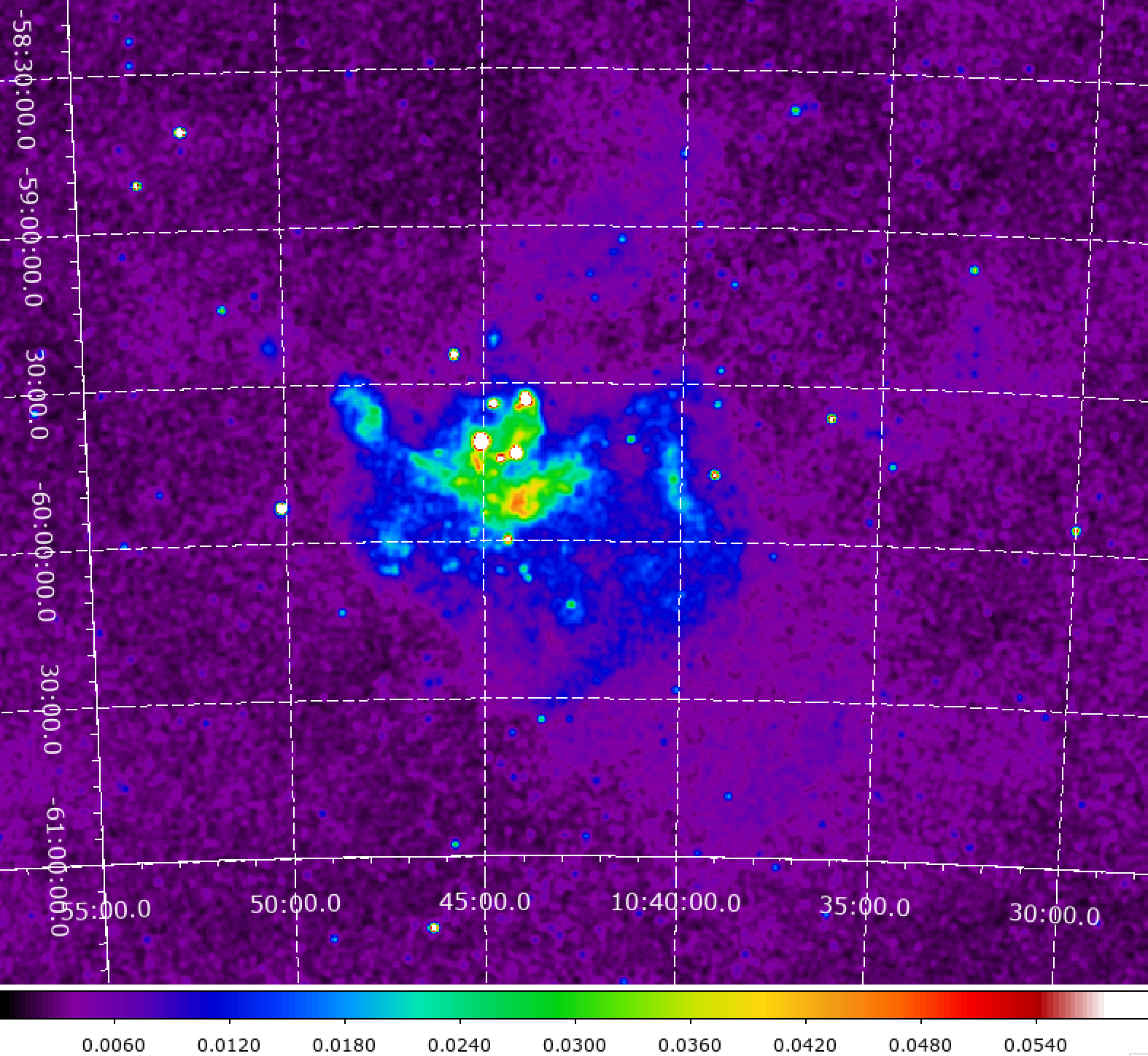}
\end{center}
\caption{\label{eROSITAima}
Exposure-corrected eROSITA broad band image (0.2 -- 10.0 keV, 0 -- 0.06 cts/s) in equatorial coordinates. The image is shown using a linear intensity scale.}
\end{figure*}

\begin{figure*}
\begin{center}
\includegraphics[width=\textwidth,trim=108 5 108 0,clip]{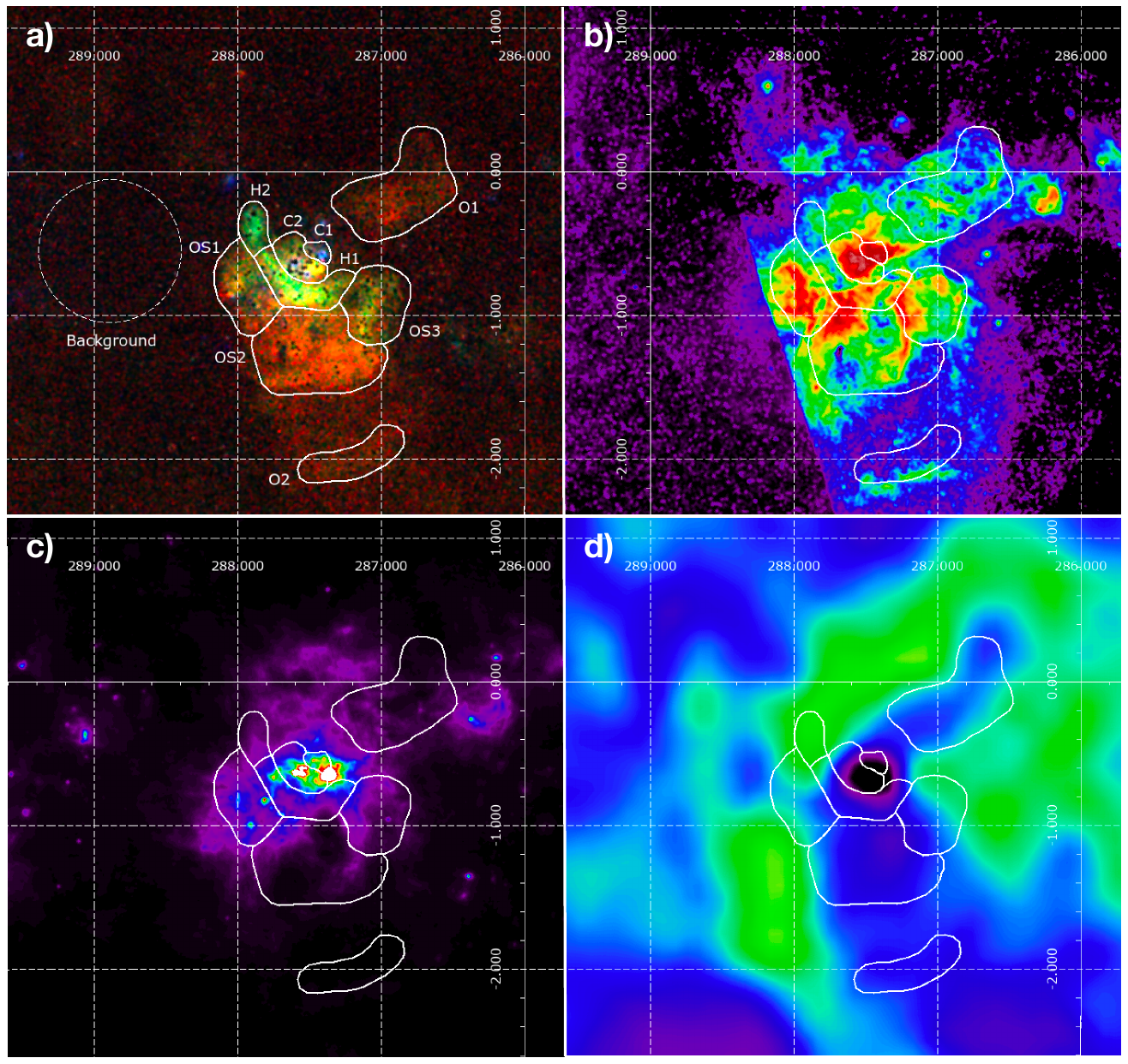}
\end{center}
\caption{\label{images}
Exposure-corrected eROSITA 
three-colour image using point-source-free images of the three softer bands (red: 
0.2 -- 0.5 keV, green: 0.5 -- 1.0 keV, blue: 1.0 -- 2.0 keV, a)),
optical $R$ band image showing the mainly photoionised gas (2nd Digitized Sky Survey, $R$ image, 6000 -- 26000 [arbitrary units], b)), 
IR image of cold dust with \akari\ in the 60$\mu$m band \citep[][0 -- 15000 MJy/sr, c)]{2015PASJ...67...50D}, 
and HI map of HI4PI survey \citep[][0 -- 1500, d)]{2016A&A...594A.116H} with extraction regions for the spectral 
analysis shown in white. 
The labels correspond to C1: Central region 1, C2: Central region 2, H1: Hook region 1, H2: Hook region 2, OS1: Outskirt region 1, OS2: Outskirt region 2, OS3: Outskirt region 3, O1: Outer region 1, O2: Outer region 2. The dashed circle shows the region where the spectrum of the local background was extracted.
All images shown in Galactic coordinates using a linear intensity scale.}
\end{figure*}

Using the energy calibrated data of eRASS:4, we  created images in the
energy bands of 0.2 -- 10.0 keV, 0.2 -- 0.5 keV, 0.5 -- 1.0 keV, 1.0 -- 2.0 keV,
and 2.0 -- 10.0 keV. The data are stored in partly overlapping sky tiles, each 
of which has a size of $3.6^\circ \times 3.6^\circ$.
First, we merged the event files of the sky tiles and recalculated the
sky coordinates with the CNC in the centre (RA = 10:43:72.8, Dec = --59:49:48).
Images were created for each energy band using a bin size of 400, 
i.e., one image pixel has a size of 20\arcsec $\times$ 20\arcsec.
Based on the images, vignetting-corrected exposure maps and exposure-corrected count-rate images were created. The average exposure time is $\sim620$ s.
In addition, we created event files and images without point sources 
(called `point-source-free' data hereafter).
We excluded all significantly detected point sources from the data. 
We use the standard eRASS point source catalogue of eRASS:4 produced by the consortium with a circular extraction region with a radius of 3 times the half-energy width (HEW) of 14.5\arcsec\ \citep[][for the catalogue of the first all-sky survey eRASS1]{Merloni23}.
In the $\sim$10 degr$^2$ field around the CNC, we excluded $\sim3000$ sources.
For $\eta$ Car, which is the brightest X-ray source in the CNC, we used a circular extraction region with a radius of 64\arcsec, i.e., 3 times larger than the HEW of eROSITA at higher energies \citep[29\arcsec\ below 3 keV and 42\arcsec\ at 4.5 keV,][]{2020SPIE11444E..4QD}, to remove as much emission from this source as possible.
Using these
filtered data, we binned images again with a bin size of 400 and created exposure maps and count-rate images in the same energy bands as for the complete data.
The count-rate image in the full band (0.2 -- 10.0 keV) is shown in Fig.\,\ref{eROSITAima}, while a three-colour image
of the point-source-free data in the bands 0.2 -- 0.5 keV, 0.5 -- 1.0 keV, and 1.0 -- 2.0 keV is shown in Fig.\,\ref{images}a with extraction regions 
for the spectral analysis (Sect.\,\ref{ana:spec}).

As can be seen in the eROSITA images, there is significant diffuse emission
in the central region, where the open clusters Trumpler 14, 16, and 
Collinder 232 as well as $\eta$ Car and the Wolf-Rayet star WR25 are located.
Around the massive stars, there is significant emission above 1 keV (appearing blue
in the three-colour image in  Fig.\,\ref{images}a in the Central region 1 (C1) around Trumpler14), 
indicating emission from the stellar winds. 
The emission at the position of the stellar cluster Trumpler 15 is also 
bright above 1 keV. 

There is bright and soft extended X-ray emission south of Trumpler 14, which extends  to the east, beyond 
Trumpler 16 and Collinder 232 (Central region 2, C2). 
The Carina nebula is very bright in this region
in the optical (see Fig.\,\ref{images}b) and in the radio continuum \citep[][see Sect.\,\ref{disc:mwl}]{2021ApJ...909...93R}. The gas seems to be photoionised strongly from
the radiation of the massive stars of the stellar clusters, $\eta$ Car, and
WR25.

The most pronounced structure in X-rays is the `X-ray hook'. It appears 
green in the three-colour image, indicating that the emission is dominated
around 1 keV, with more contribution of emission above 1 keV than in the
adjacent C2 region. 

Next to the X-ray hook (Hook region 1 and 2, H1 and H2, respectively), there is emission mainly between 0.5 keV and 1.0 keV,
indicating hot thermal plasma (Outskirt regions 1 and 3, OS1 and OS3).
In these region, shell-like structures are seen in the photoionised gas
(Fig.\,\ref{images}b), traced also by dust 
(Fig.\,\ref{images}c).

Towards lower Galactic latitudes, the emission is soft (around 0.5 keV) and 
rather faint and diffuse (Outskirt region 2, OS2, and further out, being a little brighter
in Outer region 2, O2). 
Similar emission is also observed to the north (Outer region 1, O1). 
As can be seen in the HI map (Fig.\,\ref{images}d), soft faint emission
seems to indicate hot gas escaping into cavities in the cold gas distribution.

\subsection{Spectra}\label{ana:spec}

\begin{figure*}
\begin{center}
\includegraphics[width=.47\textwidth,trim=20 10 130 50,clip]{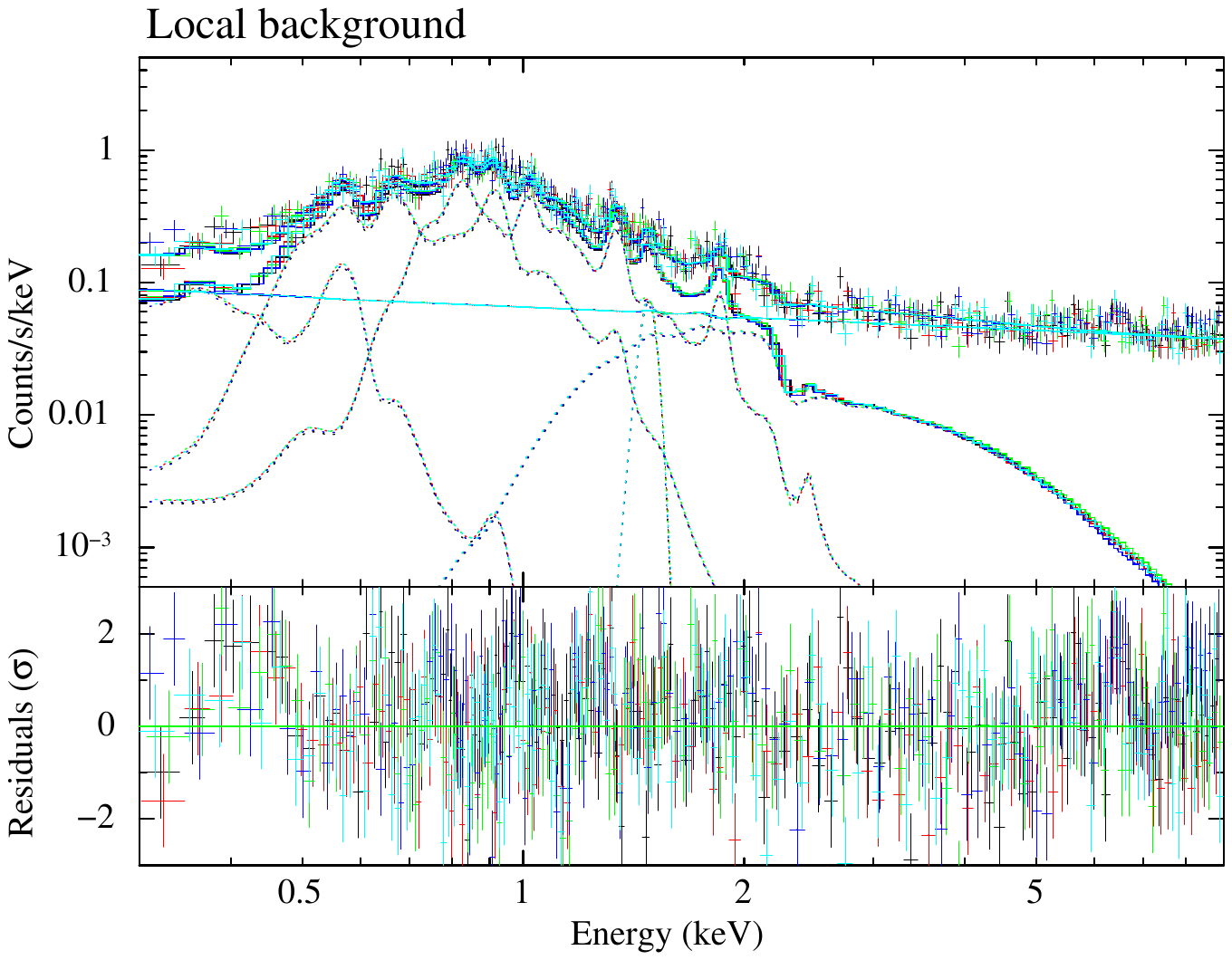}\\
\includegraphics[width=.47\textwidth,trim=20 10 130 50,clip]{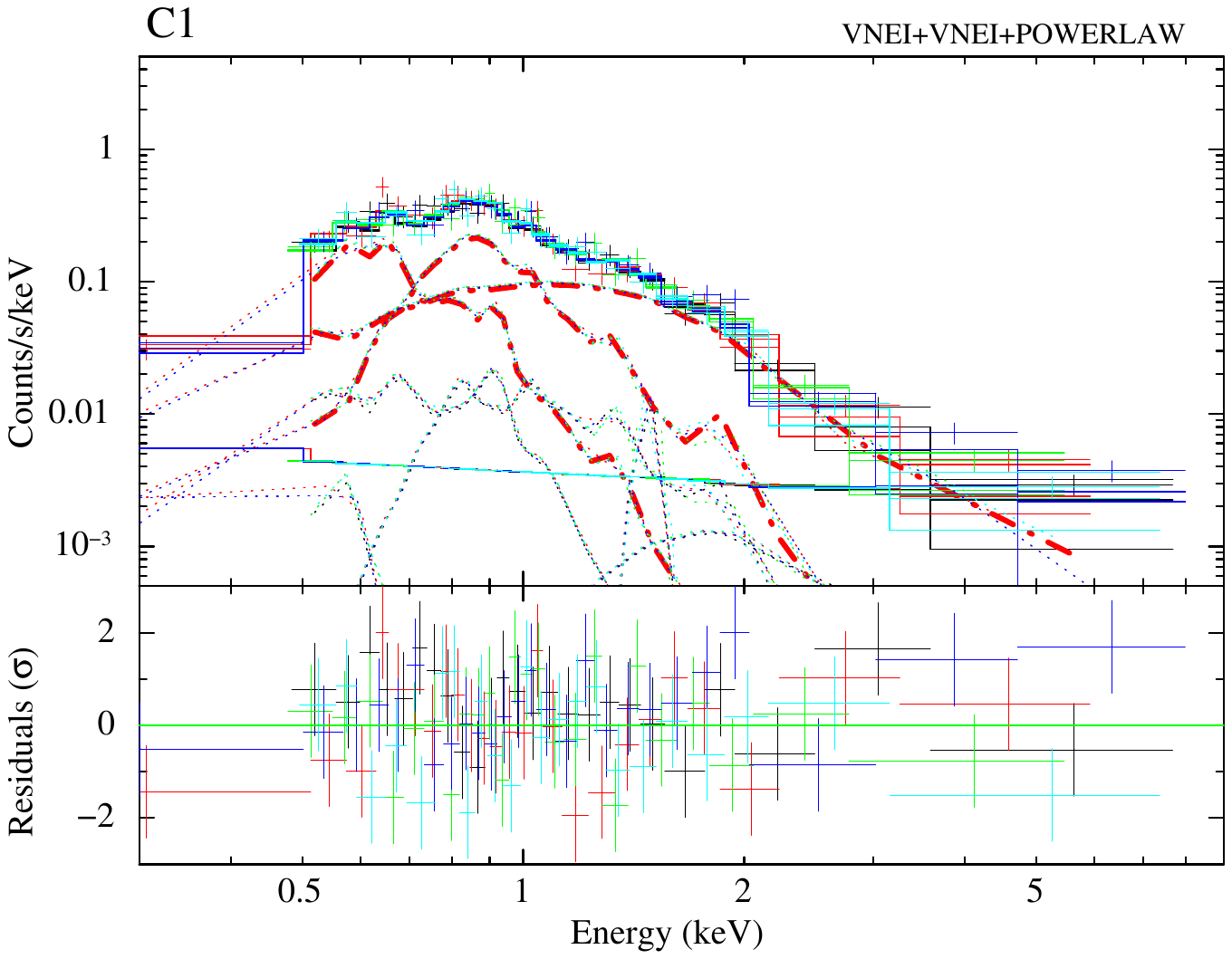}
\includegraphics[width=.47\textwidth,trim=20 10 130 50,clip]{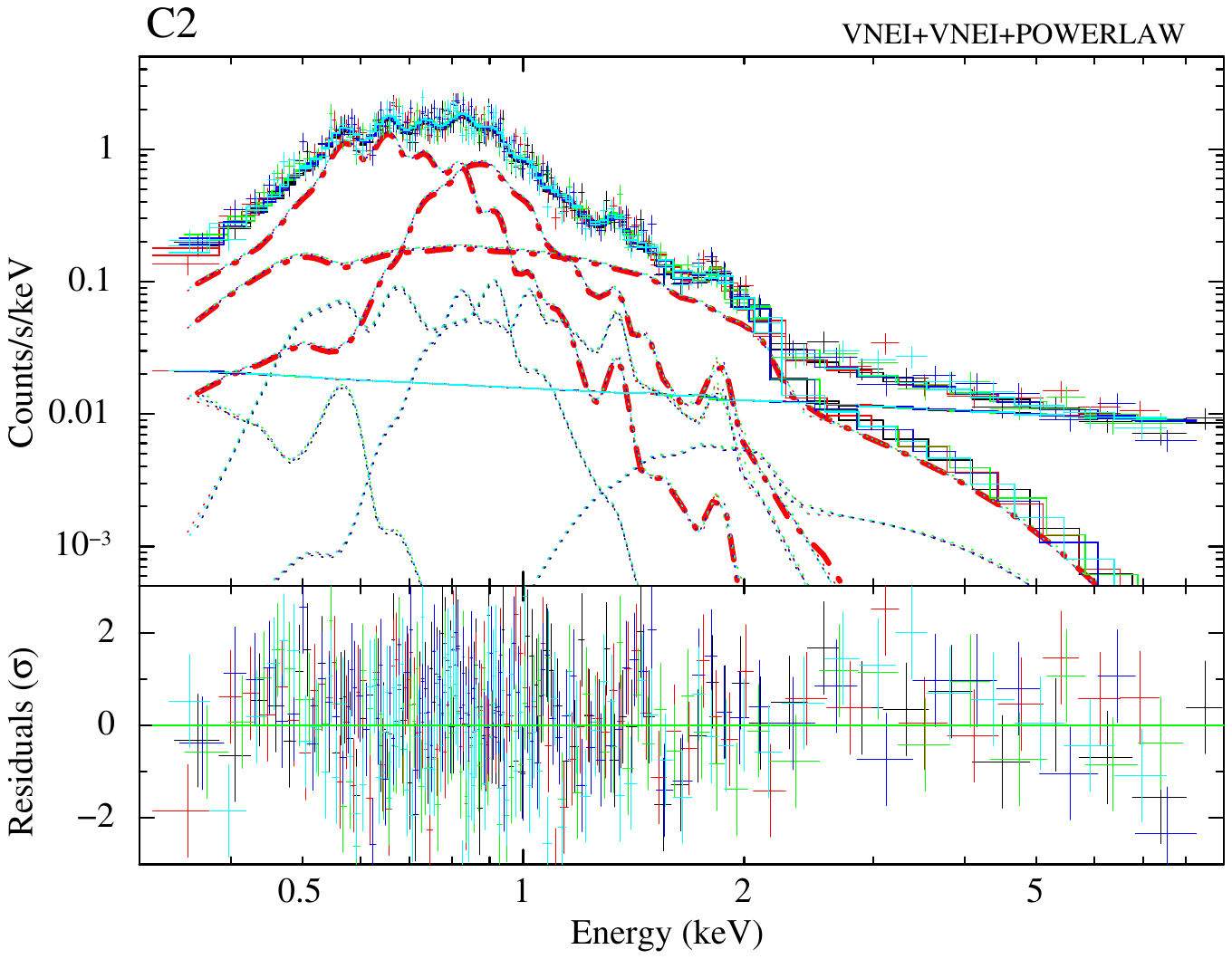}\\
\includegraphics[width=.47\textwidth,trim=20 10 130 50,clip]{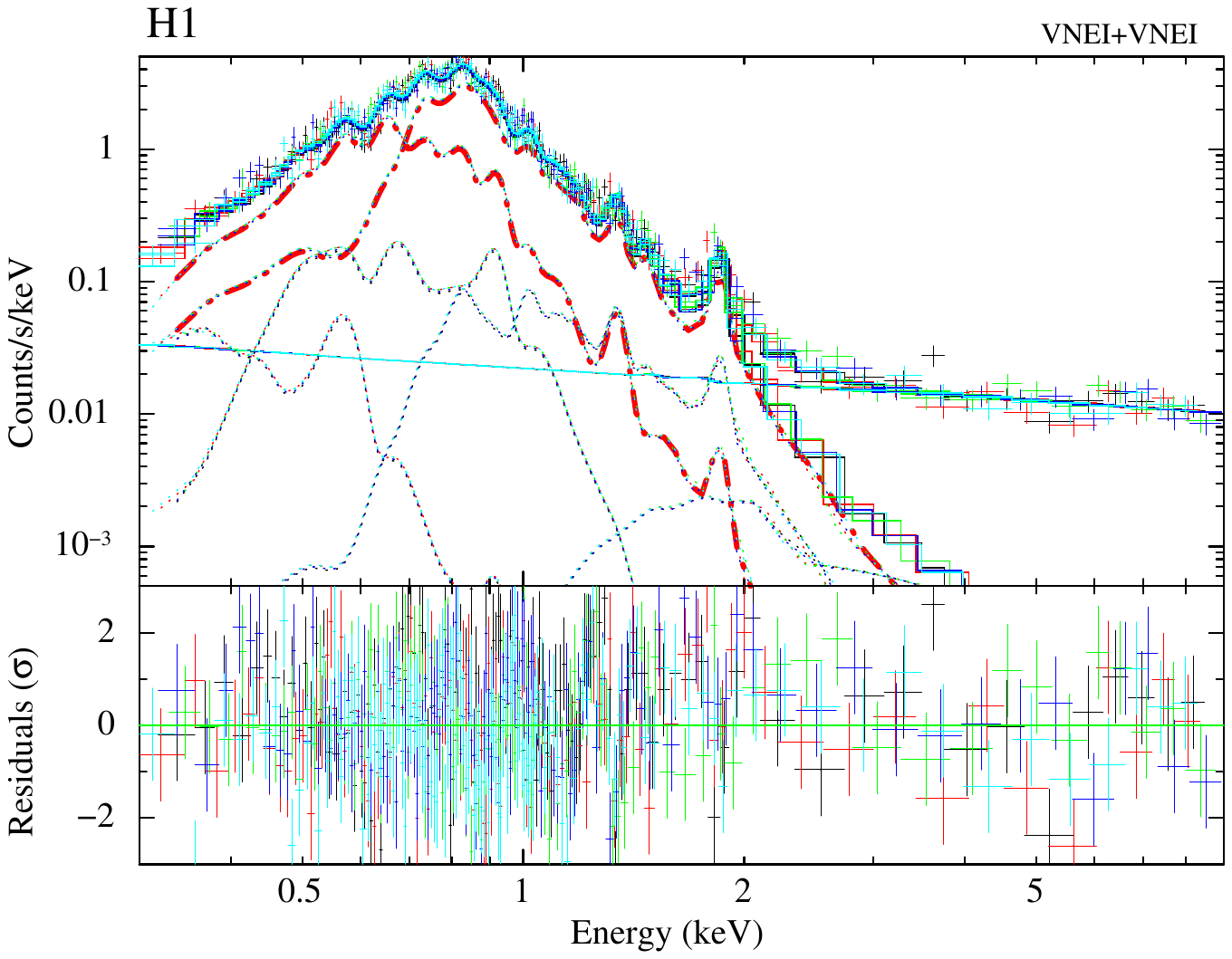}
\includegraphics[width=.47\textwidth,trim=20 10 130 50,clip]{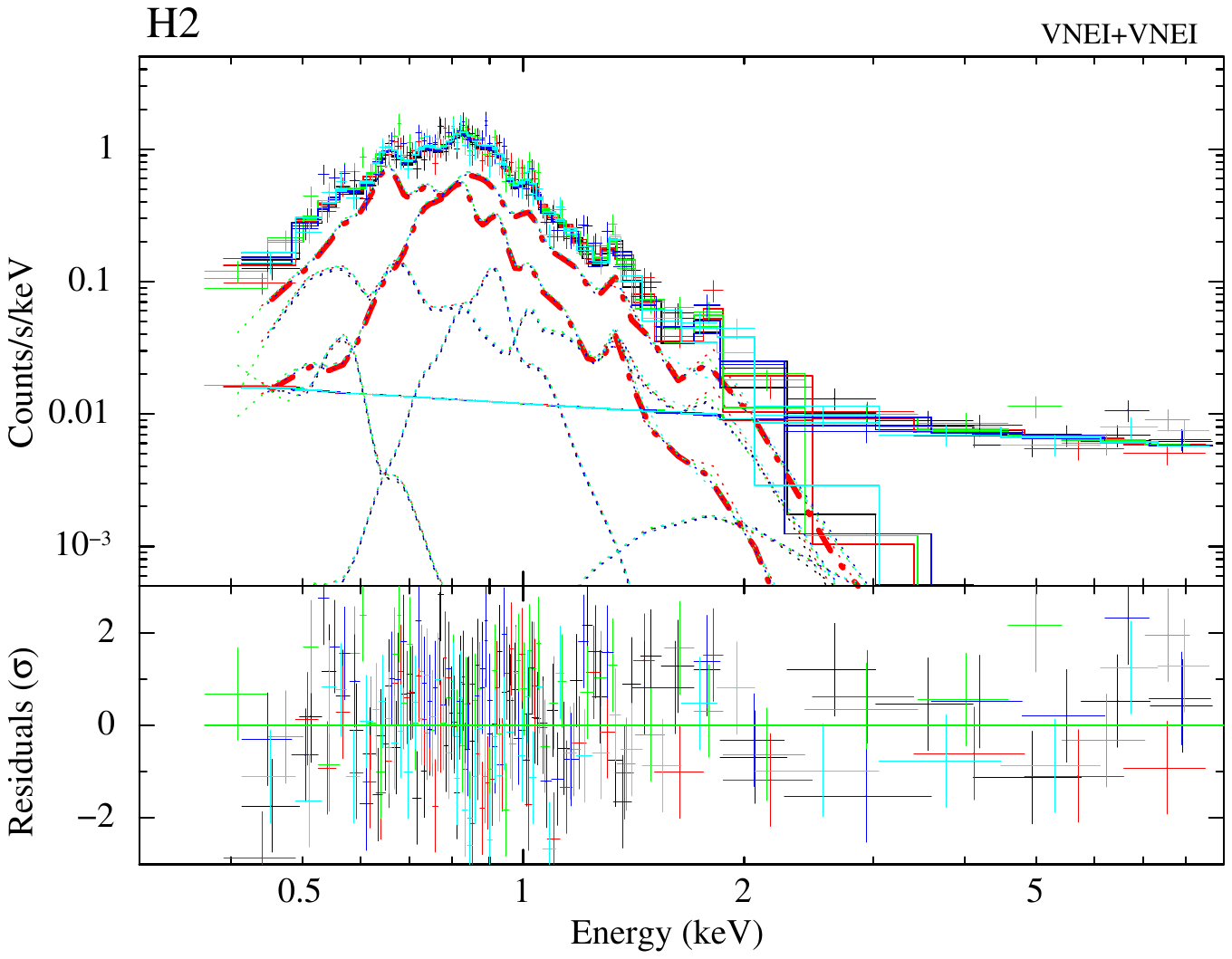}
\end{center}
\caption{\label{spectra}
eROSITA TM1-4,6 spectra are shown in the colours black, red, green, blue, cyan, respectively. For each TM we show the spectral model components with the corresponding colours (the cyan lines are best visible) with dotted lines. The model also includes the particle background (seen as a straight line) and local X-ray background components. In addition, we mark the source components with thick, red, dash-dotted lines. The extraction regions are indicated in the upper left corner of each panel.}
\end{figure*}

\addtocounter{figure}{-1}
\begin{figure*}
\begin{center}
\includegraphics[width=.47\textwidth,trim=20 10 130 50,clip]{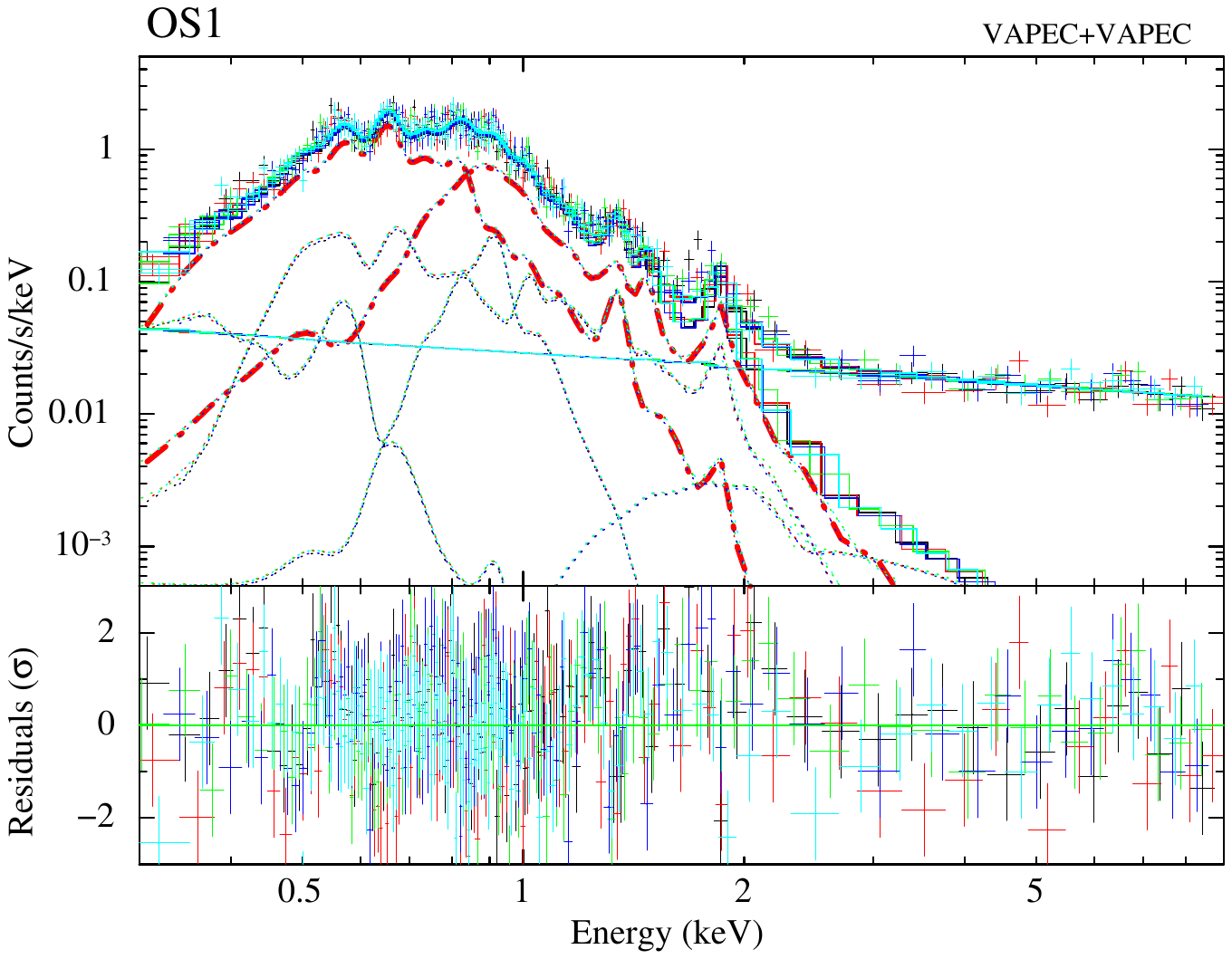}
\includegraphics[width=.47\textwidth,trim=20 10 130 50,clip]{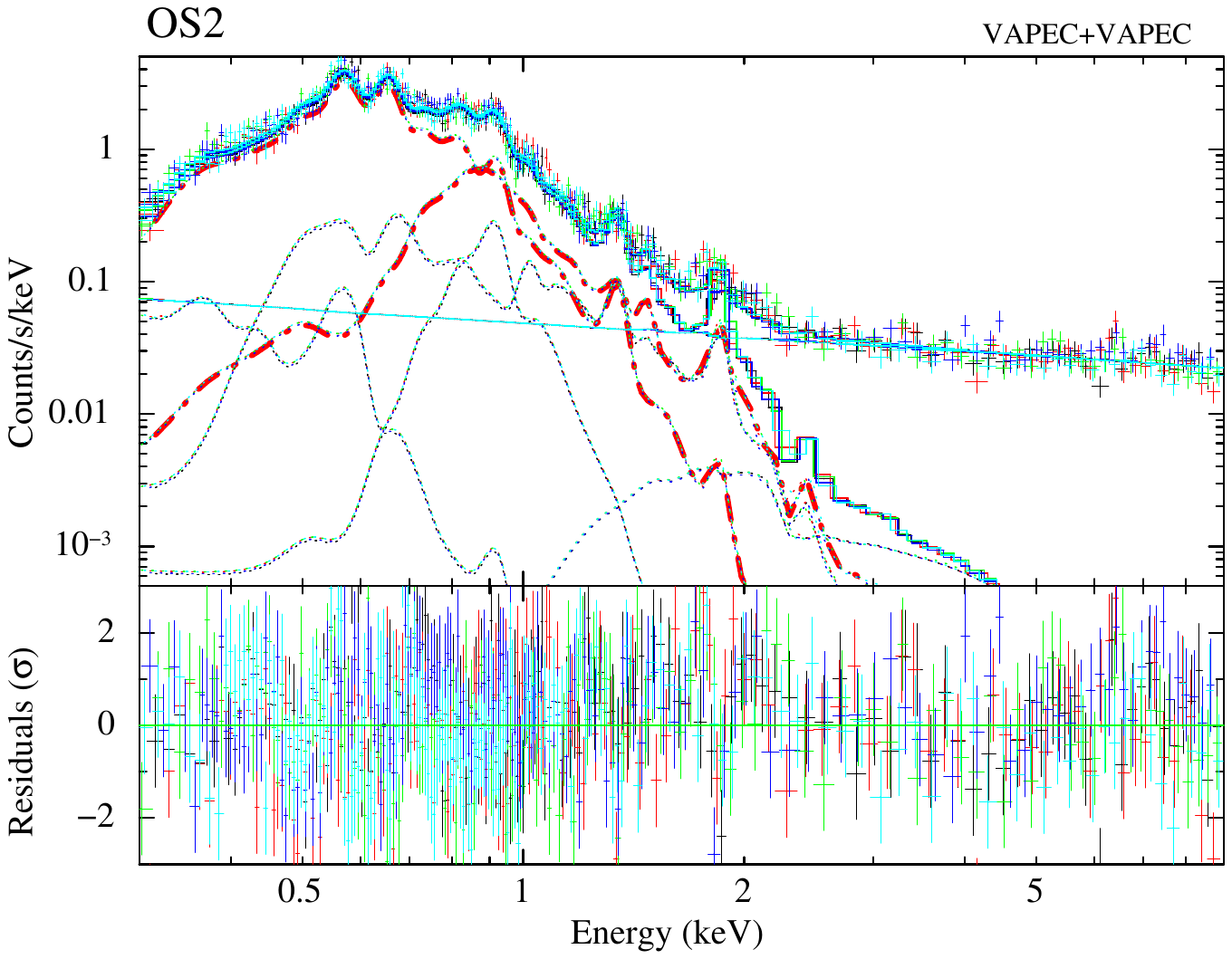}
\includegraphics[width=.47\textwidth,trim=20 10 130 50,clip]{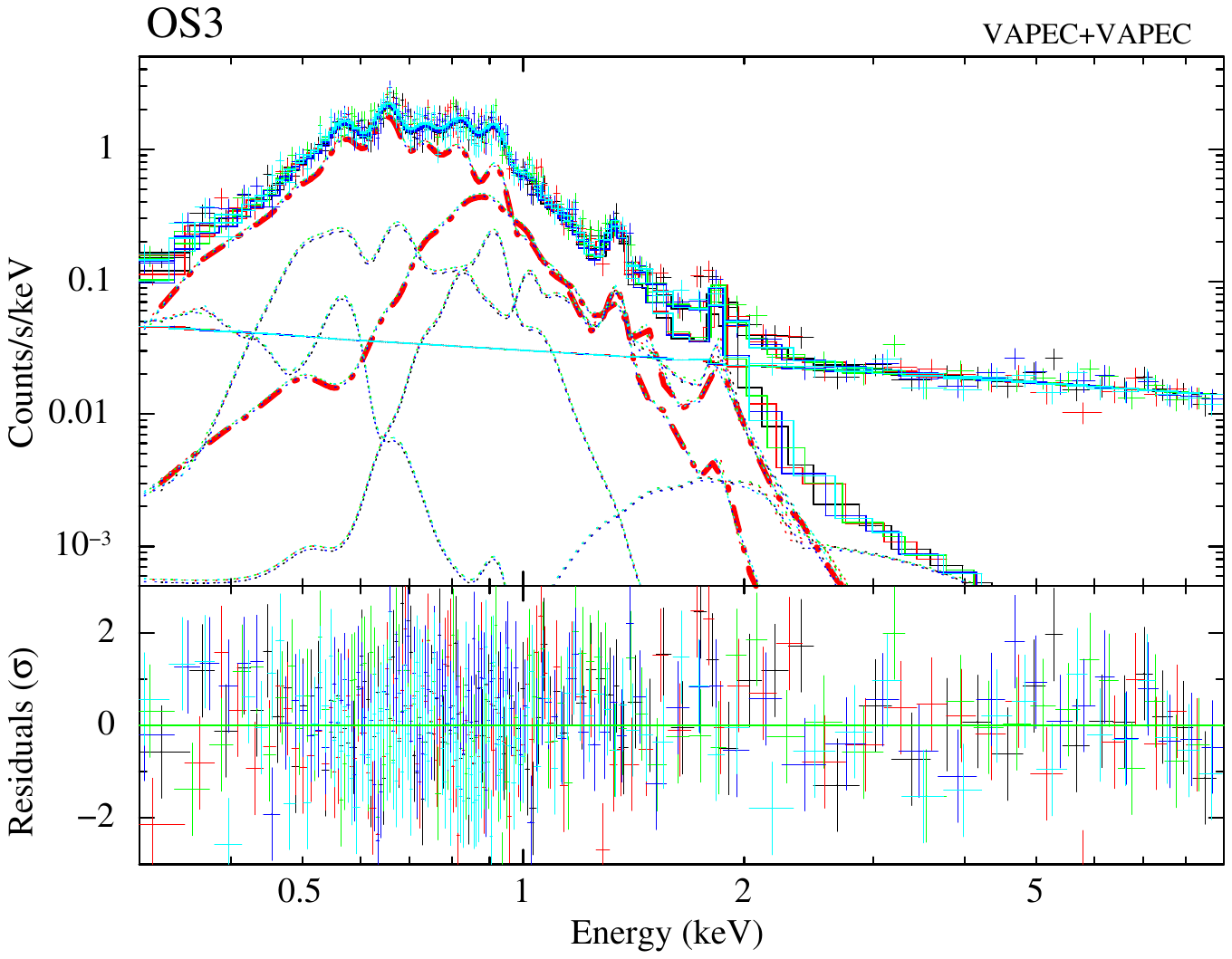}\\
\includegraphics[width=.47\textwidth,trim=20 10 130 50,clip]{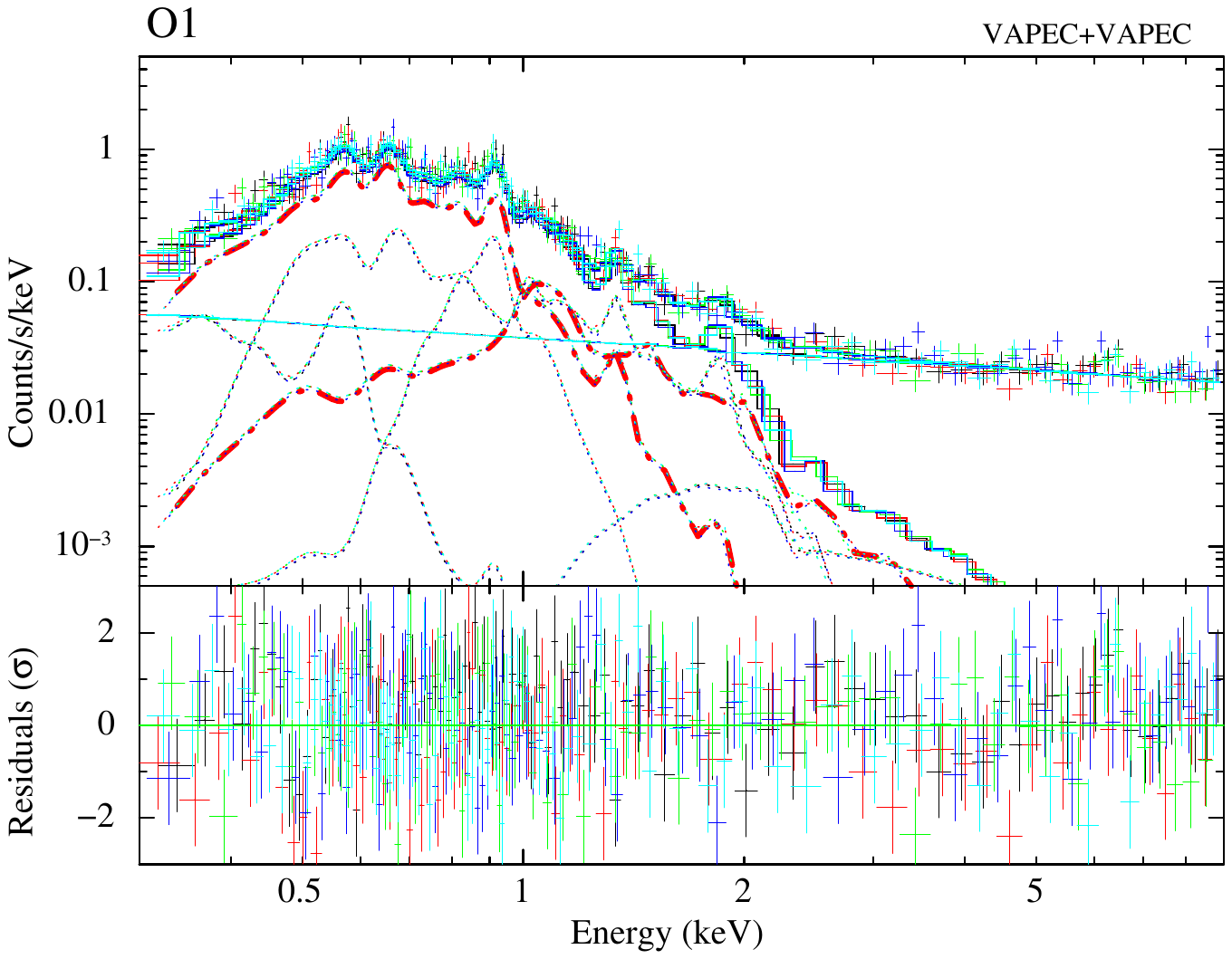}
\includegraphics[width=.47\textwidth,trim=20 10 130 50,clip]{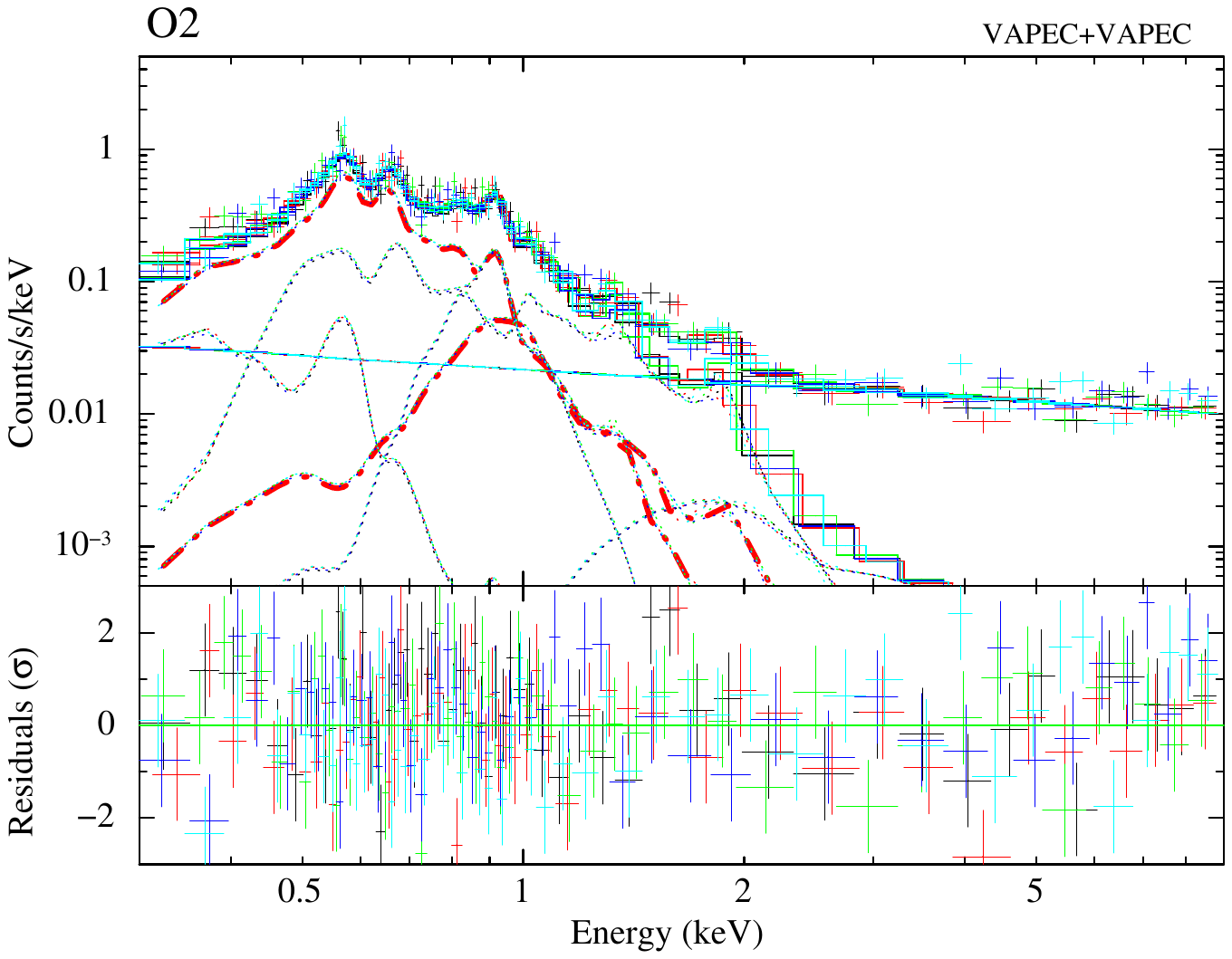}\\
\end{center}
\caption{\label{spectra2}
Continued.}
\end{figure*}

\begin{table*}
\caption{\label{spectab}
Spectral fit parameters. 90\% confidence interval is given in brackets.}
\small
\begin{tabular}{lllll}
\hline
& \multicolumn{2}{c}{Central regions} & \multicolumn{2}{c}{Hook regions} \\
\hline
%region &                               Center1 reg8 &              Center2 reg3 &              Hook1 reg4 &                Hook2 reg9  \\
region &                               C1      &              C2      &              H1      &                H2       \\
model &                                2\vnei+\pow &               2\vnei+\pow &               2\vnei &                    2\vnei \\
\hline
%21
$N_\mathrm{H}$ [$10^{22}$ cm$^{-2}$] & 0.28 (0.21--0.35) &       0.13 (0.10--0.17) &       0.13 (0.11--0.17) &       0.25 (0.20--0.30) \\ 
\hline
%24 for reg 3, otherwise 22 
$kT_1$ [keV] &                         0.24 (0.15--0.25) &       0.28 (0.23--0.39) &       0.26 (0.23--0.29) &       0.27 (0.25--0.32) \\
%38 for reg 3, otherwise 36 
$\tau_1$ [cm$^{-3}$ s] &               2.4e+11 ($>$1.7e+10) &        1.8e+11 (6.2e+10--3.9e+11) & 2.7e+11 (1.7e+11--5.4e+11) & 6.5e+11 (2.5e+11--1.3e+12) \\
%40 for reg 3, otherwise 38 
$norm_1$ &                             1.8e--3 (3.7e--4--2.5e--2) & 3.7e--3 (2.0e--3--7.6e--3) & 1.3e--2 (0.8e--2--2.3e--2) & 6.2e--3 (3.2e--3--9.9e--3) \\
\hline
%41 for reg 3, otherwise 39 
$kT_2$ [keV] &                         0.82 (0.76--0.86) &       0.82 (0.76--0.86) &       0.63 (0.62--0.65) &       0.70 (0.67--0.77) \\
%55 for reg 3, otherwise 53  
$\tau_2$ [cm$^{-3}$ s] &               1.9e+11 ($>$1.3e+11) &        3.0e+11 ($>$1.3e+11) &        3.0e+13 ($>$1.6e+12) &        3.3e+11 (2.3e+11--4.9e+11) \\
%57 for reg 3, otherwise 55  
$norm_2$ &                             6.4e--4 (3.6e--4--2.0e--3) & 1.3e--3 (1.0e--3--2.1e--3) & 1.3e--2 (1.1e--3--1.4e--3) & 2.6e--3 (1.9e--3--3.2e--3) \\
\hline
$norm_2$/$norm_1$ &                    0.35 &                      0.35 &                      1.0 &                       0.42 \\
\hline
%22 for reg 3, otherwise 56  
$\Gamma$ &                             1.7 (1.3--2.2) &          1.8 (1.5--2.3) &                                  &                           \\
%23 for reg 3, otherwise 57 
$norm_p$ &                             1.3e--3 (1.0e--3--1.8e--3) & 2.3e--3 (1.7e--3--3.0e--3) &                           &                          \\
\hline
$\chi^2$ &                             111.36 &                    513.73 &                    872.00 &                    323.29 \\
DOF &                                  126 &                       468 &                       661 &                       256  \\
\hline
$F_\mathrm{abs}^\dagger$ [erg s$^{-1}$ cm$^{-2}$] & 8.7e--12  & 2.2e--11 &  3.4e--11 & 7.6e--12 \\
$F_\mathrm{unabs}^\dagger$ [erg s$^{-1}$ cm$^{-2}$] & 1.9e--11  & 4.0e--11 & 7.5e--11 & 2.8e--11 \\
\hline
$A^*$ [deg$^2$] & 0.019 & 0.076 & 0.140 & 0.061 \\
\hline
\end{tabular}

\begin{tabular}{llll}
\hline
 & \multicolumn{3}{c}{Outskirts} \\
\hline
%region &                               Outskirts1 reg5    &        Outskirts2 reg1 &           Outskirts3 reg2 \\
region &                               OS1                & OS2                & OS3        \\        
model &                                2\vapec                   & 2\vapec                   & 2\vapec             \\      
\hline
$N_\mathrm{H}$ [$10^{22}$ cm$^{-2}$] & 0.21 (0.19--0.24)       & 0.15 (0.13--0.16)       & 0.19 (0.17--0.22)   \\    
\hline
$kT_1$ [keV] &                         0.21 (0.20--0.22)       & 0.19 (0.18--0.20)       & 0.22 (0.21--0.23)     \\  
%$\tau_1$ [cm$^{-3}$ s]               &                     &             &         &                                   \\
% 37
$norm_1$ &                             4.8e--2 (3.9e--2--6.0e--2) & 1.0e--1 (9.5e--2--1.2e--1) & 2.9e--2 (2.3e--2--3.7e--2) \\
$kT_2$ [keV] &                         0.83 (0.80--0.86)       & 0.80 (0.76--0.84)       & 0.80 (0.76--0.86)       \\
%$\tau_2$ [cm$^{-3}$ s]               &                     &             &         &                                   \\
% 53
$norm_2$ &                             6.1e--3 (5.2e--3--7.0e--3) & 5.7e--3 (4.7e--3--6.6e--3) & 2.1e--3 (1.8e--3--2.5e--3) \\
\hline
$norm_2$/$norm_1$ &                    0.13 &                      0.057 &                    0.072                    \\
\hline
%$\Gamma$                               &                             &                     &                           \\
%$norm_p$                               &                         &                           &                          \\
$\chi^2$ &                             735.26                    & 1151.54                   & 586.15                    \\
DOF &                                  601                       & 898                       & 569                       \\
\hline
$F_\mathrm{abs}^\dagger$ [erg s$^{-1}$ cm$^{-2}$] & 1.8e--11  & 4.1e--11 &  1.6e--11  \\
$F_\mathrm{unabs}^\dagger$ [erg s$^{-1}$ cm$^{-2}$] & 9.5e--11  & 1.9e--10 & 7.0e--11  \\
\hline
$A^*$ [deg$^2$] & 0.187 & 0.429 & 0.195  \\
\hline
\end{tabular}

\begin{tabular}{lll}
\hline
 & \multicolumn{2}{c}{Outer regions} \\
\hline
%region &                               Outer1 reg6 &               Outer2 reg7 \\
region &                               O1                    & O2      \\
model &                                 2\vapec                  & 2\vapec \\
\hline
$N_\mathrm{H}$ [$10^{22}$ cm$^{-2}$] & 0.18 (0.16--0.20)       & 0.12 (0.09--0.15) \\
\hline
$kT_1$ [keV] &                         0.20 (0.19--0.21)       & 0.19 (0.18--0.20) \\
%$\tau_1$ [cm$^{-3}$ s]              &                           &    \\
% 37
$norm_1$ &                             3.2e--2 (2.6e--2--3.9e--2) & 9.3e--3 (6.9e--3--1.3e--2) \\
$kT_2$ [keV] &                         1.7 (1.4--2.0)          & 0.98 (0.80--1.20) \\
%$\tau_2$ [cm$^{-3}$ s]              &                            &   \\
% 53
$norm_2$ &                             2.7e--3 (2.1e--3--3.3e--3) & 2.5e--4 (1.6e--4--3.4e--4) \\
\hline
$norm_2$/$norm_1$ &                    0.084 &                     0.027 \\
\hline
%$\Gamma$                           &                             &                           \\
%$norm_p$                           &                             &                           \\
$\chi^2$ &                             535.60                    & 329.11 \\
DOF &                                  511                       & 287 \\ 
\hline
$F_\mathrm{abs}^\dagger$ [erg s$^{-1}$ cm$^{-2}$] & 1.1e--11  & 4.6e--12  \\
$F_\mathrm{unabs}^\dagger$ [erg s$^{-1}$ cm$^{-2}$] & 5.7e--11  & 1.9e--11  \\
\hline
$A^*$ [deg$^2$] & 0.343 & 0.150   \\
\hline
\end{tabular}

$^\dagger$ Absorbed and unabsorbed flux in the energy band of 0.2--10.0 keV\\
$^*$ Size of the extraction region
\end{table*}

For the spectral analysis, we used the point-source-free data.
Therefore, the emission from the massive
stars, e.g., in $\eta$ Car is not included in the diffuse emission but will 
be addressed separately (see Sect.\,\ref{stars}). For the analysis
of the diffuse emission, however, one has to take into account that 
there might be contamination by the emission of the very bright X-ray sources in areas around them. 

We first defined extraction regions based on the
photon statistics using the Voronoi tessellation algorithm
\citep{2003MNRAS.342..345C}.
However, the diffuse emission from the CNC is very faint.
Even selecting a signal-to-noise ratio as low as 40, no reasonable spectra
were obtained, which would have allowed the analysis of the emission of the Carina nebula 
above the background.

Therefore, we decided to define the extraction regions manually in the next
step, based on
the comparison to multi-wavelength data (see Fig.\,\ref{images} and 
Sect.\,\ref{ana:ima}). The criteria we applied are the following:
\begin{itemize}
\item contiguous emission in X-rays with sufficient photon statistics,
\item similar colour in the three-colour X-ray image indicating a similar spectrum,
\item corresponding to known astrophysical objects/structures (stellar clusters, molecular clouds, interstellar cavities),
\item well correlated with distribution observed at other wavelengths (photoionised gas seen in the optical or radio, dust, molecular clouds, atomic gas).
\end{itemize}
The extraction regions are shown in Fig.\,\ref{images}. 
The region from which the spectrum was extracted to estimate the local X-ray background is shown with a dashed circle in the upper left panel.

We analysed the spectra using {\tt XSPEC} version 12.12.1.
All spectra are contaminated with particle-induced, non-X-ray background
as well as with local astrophysical X-ray background 
\citep[see][]{2020SPIE11444E..1OF,2023A&A...676A...3Y}.
An additional spectrum was extracted in a region near but well outside the 
CNC to estimate the local X-ray background.
The background was first modelled and fit, then the source spectrum
(for regions in the CNC) was simultaneously fit with the background spectrum,
also including the background components.
The particle background can be described with a multi-component powerlaw
spectrum with emission lines arising in the instruments. 
In order to estimate the particle background we used the data up
to 9.0 keV, even though no diffuse emission was observed above $\sim$3.0 keV
in most cases (above  $\sim$5.0 keV in the C1 region).
The astrophysical X-ray background consists of emission
from the Local Hot Bubble, Galactic disc and halo, and the extragalactic X-ray
background.
The spectral model parameter values, which were obtained from the fit of the spectrum of the background region, were also used for the fit of the source spectrum, from which the background had not been subtracted. The background model components were scaled with a multiplicative constant parameter, which should mainly account for the different size of the extraction regions, but was free in the fit. For example, the  area of the Central regions C1 and C2 is 0.019 deg$^2$ and 0.076 deg$^2$, while the area of the local background region is 0.785 deg$^2$, the theoretical scaling factor thus being 0.024 and 0.096 for regions C1 and C2, respectively. The best fit yielded a scaling factor of 0.02 ($<$ 0.046) for C1 and 0.11 (0.06 -- 0.23) for C2 with the 90\% confidence limits given in brackets.

X-ray observations of the ISM in the Milky Way and 
nearby galaxies have shown that the emission is well reproduced with at least 
two thermal plasma components with different temperatures
\citep[e.g.][and references therein]{2010ApJS..188...46K,2020A&A...637A..12K}.
The lower temperature component has $\sim$0.2 keV and indicates
emission from the hot phase of the ISM in equilibrium.
The higher temperature component has temperatures $>0.5$ keV and is 
particularly significant in  regions with recent additional heating 
(\hii\ regions, superbubbles, and supernova remnants, SNRs).
To model the plasma emission we used the collisional ionisation equilibrium 
(CIE) model \vapec\footnote{\url{http://atomdb.org/}} and non-equilibrium ionisation model \vnei\ \citep[][and references therein]{2001ApJ...548..820B}.

We have also fit the element abundances, in particular for O and Fe. In all regions, the fits of the eROSITA spectra suggest that the element abundances are consistent with solar values.
At first glance this looks contradictory to the results of \citet{2011ApJS..194...15T} who found enhanced abundances of O, Ne, and Mg in a few smaller regions, and super-solar abundances of Si, S, and Fe in the CNC. However, since eRASS is a shallow survey, as our first tessellation approach has shown, the photon statistics are not sufficient to analyse small regions. Therefore, enhancements of O, Ne, or Mg seen in small regions in the deep \chandra\ data will be dominated by the emission from plasma with solar abundances in larger regions. Moreover, there is an Al-K instrumental line at 1.5 keV in the eROSITA data and the sensitivity drops significantly above 2 keV, which makes it impossible to constrain the abundances of Si, S, and Fe.

In the Central regions C1 and C2, and to some extent also in the Hook regions H1 and H2, the emission appears harder than in the other regions. \citet{2011ApJS..194...15T} have shown that this emission is well reproduced by one or two additional thermal components with higher temperatures, accounting for possible emission from, e.g., pre-main-sequence stars or $\eta$ Car. We therefore fit the spectra in these regions with additional thermal components.
An alternative scenario for the hard emission is that
particles are accelerated in shocks of stellar winds and can
give rise to a non-thermal component also in the X-ray regime. 
To verify the existence of such a component, we also included a powerlaw 
emission component (\pow) in some regions. 
\citet{2011ApJS..194...15T} also considered the existence of a non-thermal component in the diffuse emission of Carina nebula, however, did neither present nor discuss any spectral fit results including a powerlaw component.

To account for absorption, we used the model \tbvarabs\ \citep{2000ApJ...542..914W}.
The spectra of the local background and the source regions with the best-fit models are shown in Fig.\,\ref{spectra} and
the spectral fit parameters are listed in Table \ref{spectab}.

The extraction region C1 covers the emission around Trumpler 14 and Collinder 232,
which are seen as bright point-like sources and have been removed.
The diffuse emission
is hard, and the spectrum shows
significant emission above 1 keV without emission line features.
For the source spectrum, an additional emission component is necessary to reproduce the emission between 1 and 5 keV. We first added a third thermal plasma component for the source model. 
As an alternative, we also fit the spectrum including a powerlaw component instead of the third thermal component. Both the model with a third thermal component and with a non-thermal component yield similarly good fits. A model including two additional thermal components with $kT = 2.5, 6.0$ keV similar to the model of \citet{2011ApJS..194...15T} for the hard emission also yields comparable fit statistics.
If we assume one additional thermal component, the temperature becomes very high ($kT$ = 6.8 (4.7 -- 12) keV, with 90\% confidence range in brackets), rather unphysical for shocked interstellar plasma.
The spectrum is fit well with a combination of a two-component thermal plasma model assuming non-equilibrium ionisation (NEI)
with \kt\ = 0.24 keV and 0.82 keV 
and a powerlaw component with photon index $\Gamma$ = 1.7 (1.3 -- 2.2),
which can be explained by a population of non-thermal electrons.
A cavity is seen
in the HI map (Fig.\,\ref{images}d). Most likely, the
cold material around the stellar clusters has been ionised and blown away
by the radiation and the stellar winds of the massive stars in the clusters.

In the region C2, there is diffuse emission around Trumpler 16. The bright X-ray emission from the massive star binary $\eta$ Car and from the star
WR25 along with other detected point sources inside the region has been
removed. The emission appears yellow-ish in the three-colour-image (Fig.\,\ref{images}a), with dominant emission in the medium band, but 
significant contributions from the soft and the hard bands. The best-fit
model requires two thermal NEI plasma components 
(\kt\ = 0.28 keV and 0.82 keV with \net\ $\approx 10^{11}$ \cms\ for both)
and an additional harder component. In this region, the fit with a non-thermal component is preferred to a fit with two additional thermal components as in \citet{2011ApJS..194...15T} with $\chi^2$ = 513 and degrees of freedom (dof) = 468 compared to $\chi^2$ = 561 and dof = 466 for a pure thermal model. If we again assume one additional thermal component, the fit yields a very high temperature of $kT$ = 15 keV with a lower limit of 9 keV (unconstrained to higher values).
In the best-fit model, the non-thermal component has a photon index of $\Gamma$ = 1.8 (1.5 -- 2.3).
This component again suggests the existence of non-thermal electrons. 

Next to the region C2, we have defined the X-ray hook region H1, which is located
south of the triangular shaped bright region in the optical and IR, which
also in part coincides with our region C2 (Fig.\,\ref{images}b and c).
Relative to the massive star clusters in the CNC, the X-ray hook
is located outside the bright photoionised nebula, indicative 
of a shock of the expanding \hii\ region. 
The emission appears yellow-green, with dominant emission between 
0.5 -- 1.0 keV. The spectrum is well reproduced with two thermal components 
with \kt\ = 0.26 keV and 0.63 keV, with the first being consistent with NEI
(\net = 2.7 $\times 10^{11}$ \cms), while the hotter component seems to have
reached collisional ionisation equilibrium (CIE). In this region, the contribution of the hotter component is as strong as that of the cooler component ($norm_2/norm_1$ = 1), while in the other regions, it is less than half ($norm_2/norm_1 <$ 0.42) or almost negligible. Both the high ionisation timescale and the relatively high brightness (and hence high $norm_2$) for the hotter component suggest that the gas might be highly compressed. 

The tip of the X-ray hook (region H2) appears greener than the  
region H1. While the region H1 traces the edge of a cavity seen in HI
(Fig.\,\ref{images}d), the region H2 coincides with a larger HI
structure seen in the north. The spectrum of the region H2 suggests a 
similar intrinsic spectrum as in the region H1 (two thermal components with
two temperatures), while the hotter component also seems to be in NEI in 
the region H2. The foreground absorption is higher in this region 
(\nh\ = 2.5 (2.0 -- 3.0) $\times\ 10^{21}$ \cm, suggesting that the tip of
the X-ray hook is located inside or behind the HI structure. 
In both Hook regions, there is no indication for a 
non-thermal component.

The other regions, in which there is diffuse X-ray emission with sufficient
surface brightness that allows us to perform spectral analysis reasonably
well, the emission appears red to orange, with most of the emission below
1 keV (regions OS1, OS2, OS3 as well as O1 and O2). 
These regions coincide with lower-density regions in the HI map. Most likely
it is emission from hot plasma, escaping from the central active regions.
The spectra are well reproduced with two thermal components for plasma
in CIE with a lower-temperature component with \kt\ $\approx$ 0.2 keV and
a higher-temperature component with \kt\ $> 0.8$ keV. The region O2 extends  north of and along a broad optical filament, likely photoionised gas delineating the end of the cavity. The hot gas streaming through the HI cavity will likely be adiabatically compressed (hence heated) because it is being pushed against the HI cavity wall. Mixing with HI gas then  creates the more luminous, colder component. The fact that the region O2 is spatially anti-correlated with the optical region to the south suggests that the cavity elongates southwards approximately in the plane of the sky. The region O1 is also associated with optical emission, but in this case, X-rays and optical emission are more co-spatial, which suggests that the associated HI cavity is significantly inclined to the plane of the sky.

\section{Stellar sources}\label{stars}

\begin{figure}[t]
\begin{center}
\includegraphics[width=85mm]{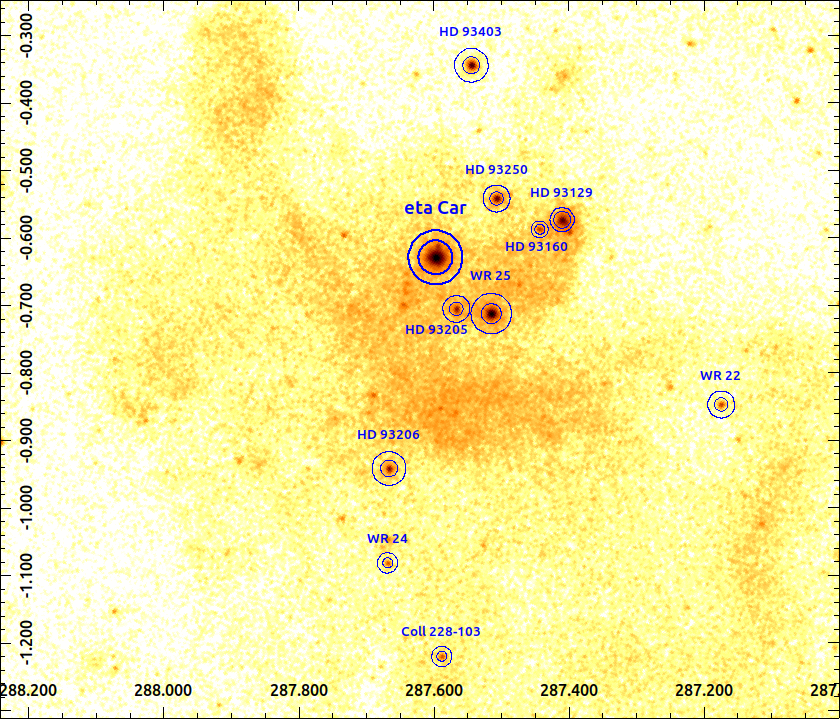}
\end{center}
\caption{\label{stars_image}eRASS:4 image and stellar sources with extraction regions in Galactic coordinates.}
\end{figure}

The Carina nebula complex contains several X-ray bright point sources associated with massive stars and stellar systems, which have been studied extensively over the last decades, see, e.g., the {\it Chandra} based CCCP \citep{2011ApJS..194....1T, 2011ApJS..194....5G, 2011ApJS..194....7N}. While the CNC harbours thousands of young stars, only the X-ray brightest ones are detected as individual sources in eRASS (see Table \ref{stars_id}). In our analysis of the eRASS data we thus focus on the prominent X-ray sources in the region, i.e., $\eta$~Car itself, several O-type stars and binaries, and the Wolf-Rayet stars WR~22, WR~24 and WR~25. An intensity scaled eRASS image with the studied stars and used source regions is shown in Fig.~\ref{stars_image}.

For the point source analysis, events from all TMs were used. Source and background photons were extracted from respectively adapted circular and annular regions. While some minor contributions from other nearby stars and diffuse emission are present, for the eRASS sources discussed here, the photon distribution is in most cases strongly dominated by the X-ray emission from a single star or a stellar system. With the exception of $\eta$~Car we use the eRASS:4 combined spectra in our analysis and thus derive the average properties of the sources. The  X-ray spectra are  fit with absorbed multi-temperature CIE plasma models (\apec) with  errors denoting the 90\% confidence range. A foreground column density of $2\times10^{21}$~cm~$^{-2}$ is used to account for the interstellar foreground absorption and the adopted distance to the CNC is 2.3~kpc.

\begin{figure}[t]
\begin{center}
\includegraphics[width=88mm]{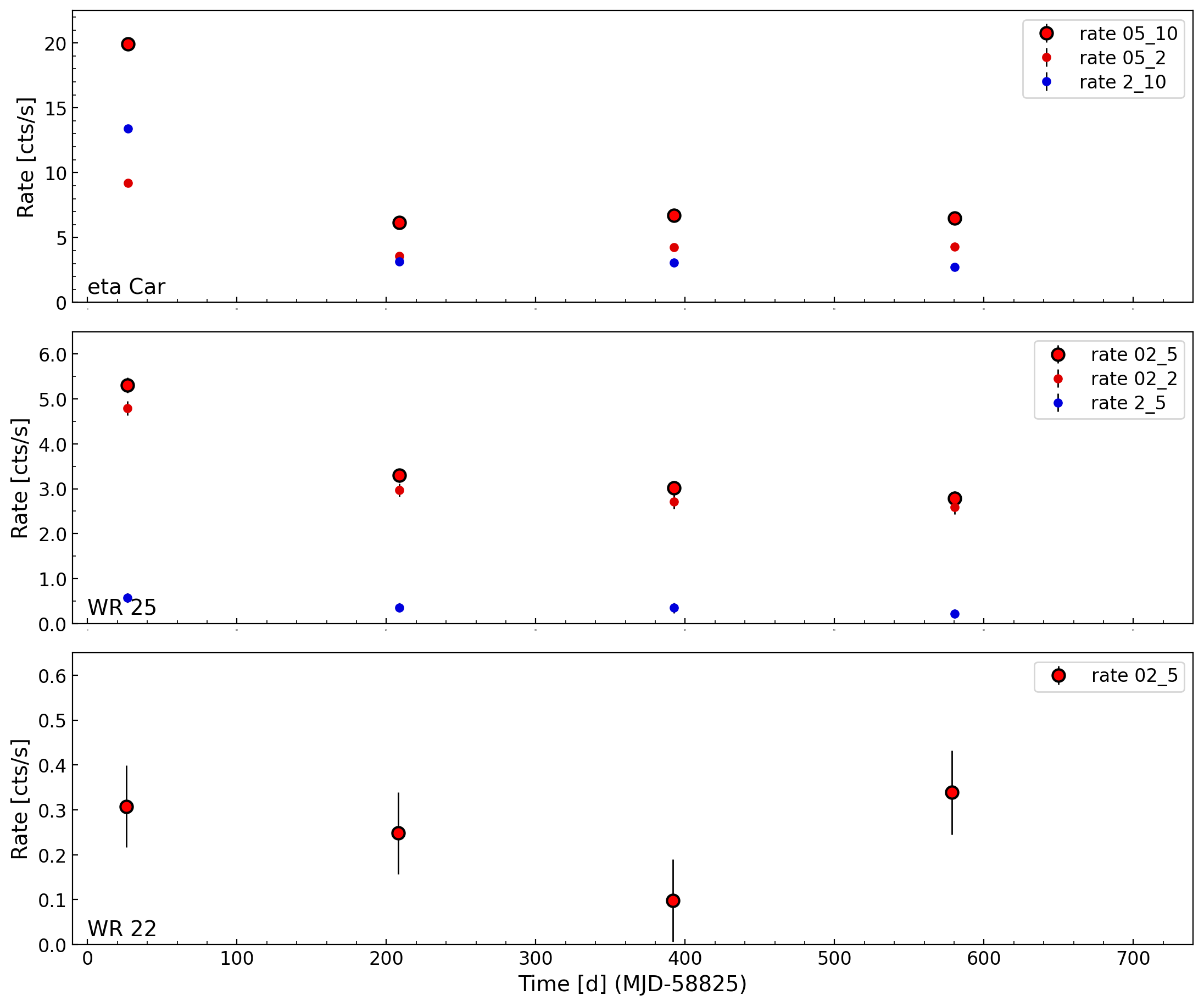}
\end{center}
\caption{\label{stars_lcs}Light curves of stellar sources in the CNC from eRASS, here highlighting the X-ray brighter targets $\eta$~Car, WR~25, and WR~22, and their  variability. The legend denotes the used energy band in keV of the respective light curve (e.g., 05\_10: 0.5 -- 10 keV).}
\end{figure}

During the eRASS mapping, the region was observed four times with a long-term sampling of about 0.5~yr. The exposure times per individual eRASS, obtained in multiple scans over 2\,--\,3 days, are around 0.5~ks and provide sufficient photons to study the X-ray variability for our targets. Example light curves of three selected X-ray sources are shown in Fig.~\ref{stars_lcs}.

\subsection{$\eta$ Carinae}

$\eta$~Carinae (HD 93308), the most prominent stellar object in the CNC, is a very massive stellar binary in a highly eccentric 5.5~yr orbit located in Trumpler~16. It consists of an eruptive primary in its luminous blue variable (LBV) phase and a secondary, potentially evolved high-mass star with mass estimates of $M_{1} \gtrsim 80-100$ and $M_{2} \gtrsim 30-40~M_{\odot}$. A discussion of the $\eta$~Car binary parameters can be found, e.g., in \cite{2001ApJ...547.1034C, 2016ApJ...825..105K}. In this system, the collision of the fast wind from the secondary and the slower but more massive wind from $\eta$~Car~A leads to the complex and variable X-ray emission and absorption scenarios over the orbit, especially around periastron \citep[see, e.g.,][]{2002A&A...383..636P, 2007ApJ...663..522H, 2017ApJ...838...45C}.

\begin{figure}[t]
\begin{center}
\includegraphics[width=88mm]{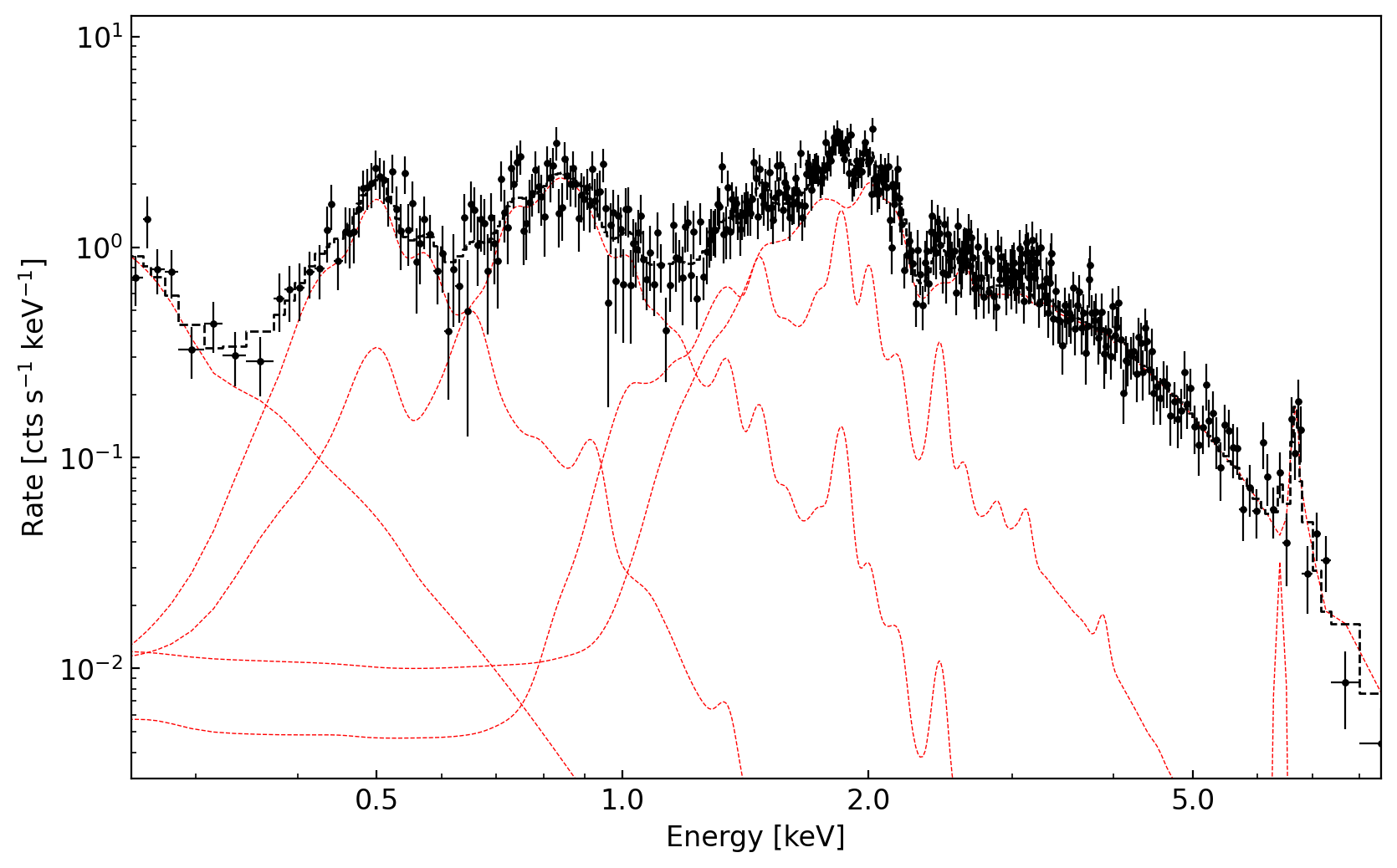}
\includegraphics[width=88mm]{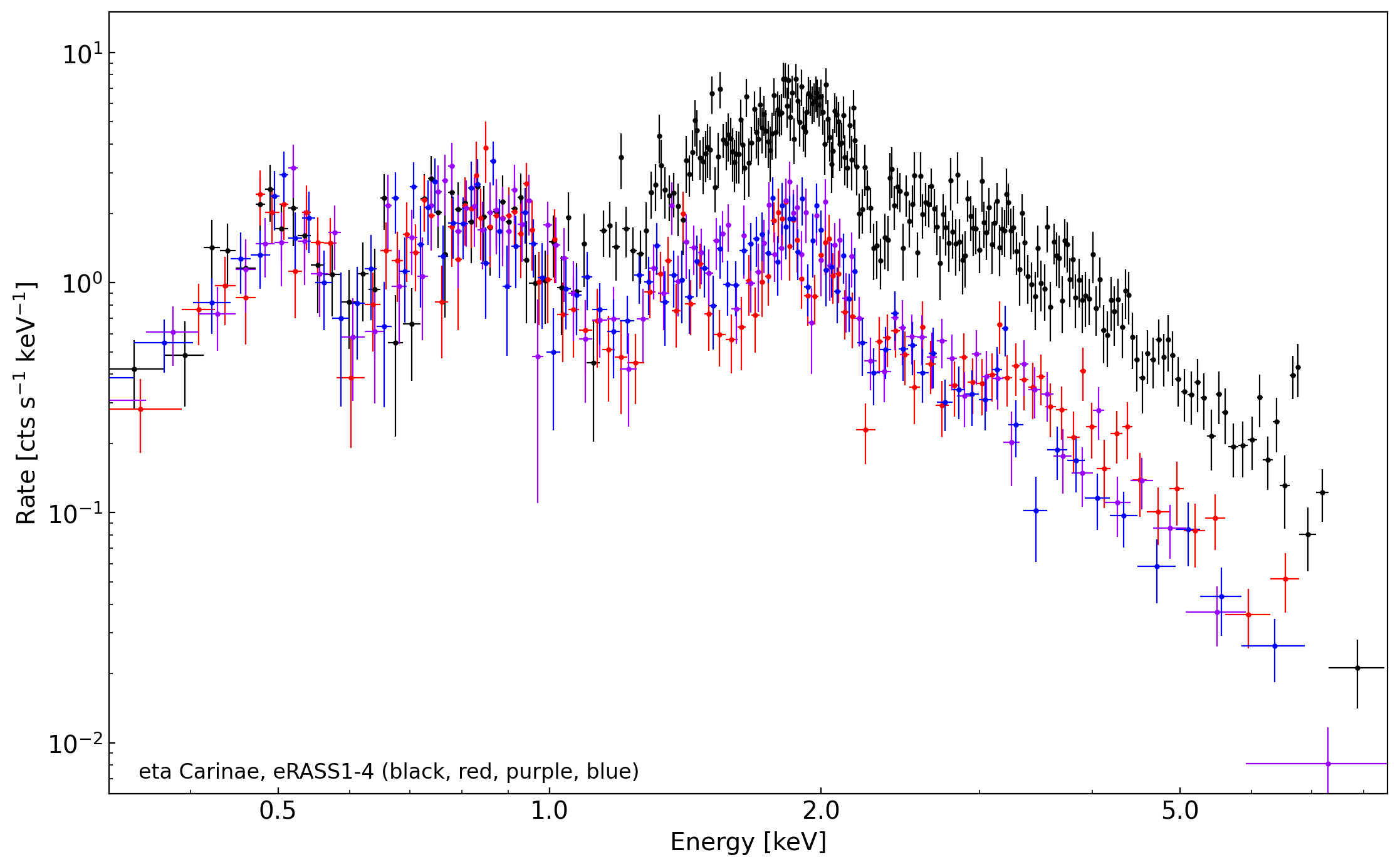}
\end{center}
\caption{\label{etaCar_spec} Spectra of $\eta$ Car as observed with eROSITA. Top: Combined eRASS:4 spectrum with spectral model (black) and model components (red), bottom: eRASS\,1-4 spectra, shown in black, red, purple, and blue.}
\end{figure}

X-rays from $\eta$~Car include emission from the outer structures created by the ejecta of the last eruptions, e.g., the famous Homunculus nebula and a more absorbed and variable hard component from the inner wind shock regions \citep{2022ApJ...937..122C}. The source is not spatially resolved by eROSITA and the data provide a combined source spectrum. Due to the optical/IR-brightness of the star ($G=4.0$~mag, $J=1.7$~mag), a mild contamination from optical loading is present at energies of $\lesssim 0.3$~keV, and is included as a background component in the modelling. The combined eRASS data are first used to derive a `baseline model' that describes the non-variable spectral components in our $\eta$~Car data. We then study the variability between the individual eRASS visits. Particularly in eRASS1, the data taken on January 02 -- 05, 2020, stand out, as this observation covers the X-ray bright state before the periastron passage. The obtained spectra are shown in Fig.~\ref{etaCar_spec}. 

The X-ray emission from the stable component, associated to the outer ejecta/nebula around $\eta$ Car, is well described by a two-temperature plasma with components at $kT_{1,2} = 0.15$ keV and 0.7~keV, with the ratio of the emission measures $EM1/EM2 \approx 1$, a high N overabundance by a factor of 15 and an absorption column fixed at the ISM value of $N_{\rm H}=2\times10^{21}$~cm$^{-2}$. From the observed $F_{\rm X} = 1.8 \times 10^{-12}$~erg\,cm$^{-2}$\,s$^{-1}$ we derive an emitted 0.2\,--\,10.0~keV X-ray luminosity of $3.7 \times 10^{33}$~erg\,s$^{-1}$ for the emission from the surrounding region. The outer ejecta is well described by a constant emission component in our data, albeit a slow temporal evolution is expected for the expanding nebula. Comparing our eRASS results with older X-ray measurements, we find that the derived flux from the combined eRASS 2020/21 data matches the long-term trend seen in X-rays, which indicates a $t^{-3}$ flux decline, as deduced from a fit to \xmm\ EPIC-pn data from observations performed between 2003 and 2015 \citep{2022ApJ...937..122C}. For the  0.3\,--\,1.0~keV energy band we get $F_{\rm X} = 1.3 \pm 0.1 \times 10^{-12}$~erg\,cm~$^{-2}$\,s~$^{-1}$ from our spectral model, well matching the fitted X-ray trend over two decades and supporting their Great Eruption and blast wave evolution scenario.

Further, a stronger absorbed, variable plasma component is detected. It is thought to be associated with the wind-wind-collision zone and one or two hot plasma components are required to model the eRASS spectra. Being most prominent during the bright state, Fe~K$\alpha$ 6.4~keV emission is detected, e.g., with a line flux of $4.7 \times 10^{-4}$~photons\,s$^{-1}$ in eRASS1. To put our data into context, we use the orbital X-ray phasing, MJD\_min = $50799.4 + 2023.4 \times n$, following \cite{2017ApJ...838...45C}.

In total, the mean eRASS X-ray luminosity of $\eta$~Car in the 0.2\,--\,5.0 + 5.0\,--\,10.0 keV band is $(2.8 + 2.9) \times 10^{34}$~erg\,s$^{-1}$ (observed), $(3.1 + 2.9) \times 10^{34}$~erg\,s$^{-1}$ (ISM $N_{\rm H}$ corrected) and $(13.0 + 3.1) \times 10^{34}$~erg\,s$^{-1}$ (full $N_{\rm H}$ corrected). The eRASS1-4 spectral modelling results for $\eta$~Car are summarised in Tab.~\ref{etacar_s}.

\begin{table*}[t]
\caption{\label{etacar_s}Spectral model for $\eta$ Car, inner wind-wind collision zone emission. Observed $F_{\rm X}$ is the total source flux in the 0.2-10.0/2-10 keV band.}
\begin{center}
\begin{tabular}{rrrrrrrrr}\hline\hline\\[-3mm] 
phase & $F_{\rm X~obs}$  & $N_{\rm H~1}$ & $kT_{1}$ & $EM_{1}$  & $N_{\rm H~2}$ & $kT_{2}$ & $EM_{2}$ & $\chi ^{2}_{\rm red}$ (dof)\\
   & $10^{-11}$ erg\,cm~$^{-2}$\,s~$^{-1}$ & $10^{22}$ cm~$^{-2}$ & keV & $10^{57}$\,cm~$^{-3}$ & $10^{22}$ cm~$^{-2}$ & keV & $10^{57}$\,cm~$^{-3}$ & \\\hline\\[-3mm]
0.979 & 20.8/19.8 & 3.10 $^{ +1.26}_{ -0.54}$ & 0.54 $^{ +0.33}_{ -0.28}$ & 0.49 $^{ +9.94}_{ -0.29}$ & 3.79 $^{ +0.35}_{ -0.31}$ & 5.36 $^{ +1.23}_{ -0.87}$ & 1.50 $^{ +0.17}_{ -0.14}$ & 1.04 (264)\\\\[-3mm]
0.069 & 4.9/4.5 & 3.28 $^{ +0.53}_{ -0.51}$ & 0.99 $^{ +0.17}_{ -0.17}$ & 0.24 $^{ +0.11}_{ -0.09}$ & 9.84 $^{ +4.61}_{ -2.71}$ & 2.75 $^{ +1.76}_{  -0.86}$ &  0.85 $^{ +0.89}_{ -0.35}$ & 0.95 (109)\\\\[-3mm]
0.160 & 3.9/3.4 & 3.5 & 0.53 $^{+0.67}_{-0.22}$ & 0.25 $^{+0.59}_{-0.15}$ & 3.5 & 3.63 $^{+4.01}_{-0.68}$ &  0.34 $^{+0.05}_{-0.13}$ & 1.02 (98)\\\\[-3mm]
0.253 & 3.6/3.1 & 3.1 & 0.96 $^{+0.55}_{-0.59}$ & 0.12 $^{+0.08}_{-0.06}$ & 3.1 & 4.32 $^{+9.57}_{-1.38}$ & 0.25 $^{+0.08}_{-0.10}$ & 0.92 (98)\\\hline
\end{tabular}
\end{center}
\tablefoot{The $\eta$~Car spectrum includes emission from the surrounding nebula, described by an 2-T \vapec\ model: $N_{\rm H}=2\times 10^{21}$cm~$^{-2}$, $kT_{1} = 0.15$~keV, $kT_{2}=0.7$~keV, $\log EM_{1} =55.80$~[cm$^{-3}$], $\log EM_{2}=55.76$~[cm$^{-3}$], N = 15 $\times$ solar, else solar abundances. For eRASS 3+4 a single $N_{\rm H}$ was used.}
\end{table*}

\begin{figure}[t]
\begin{center}
\includegraphics[width=88mm]{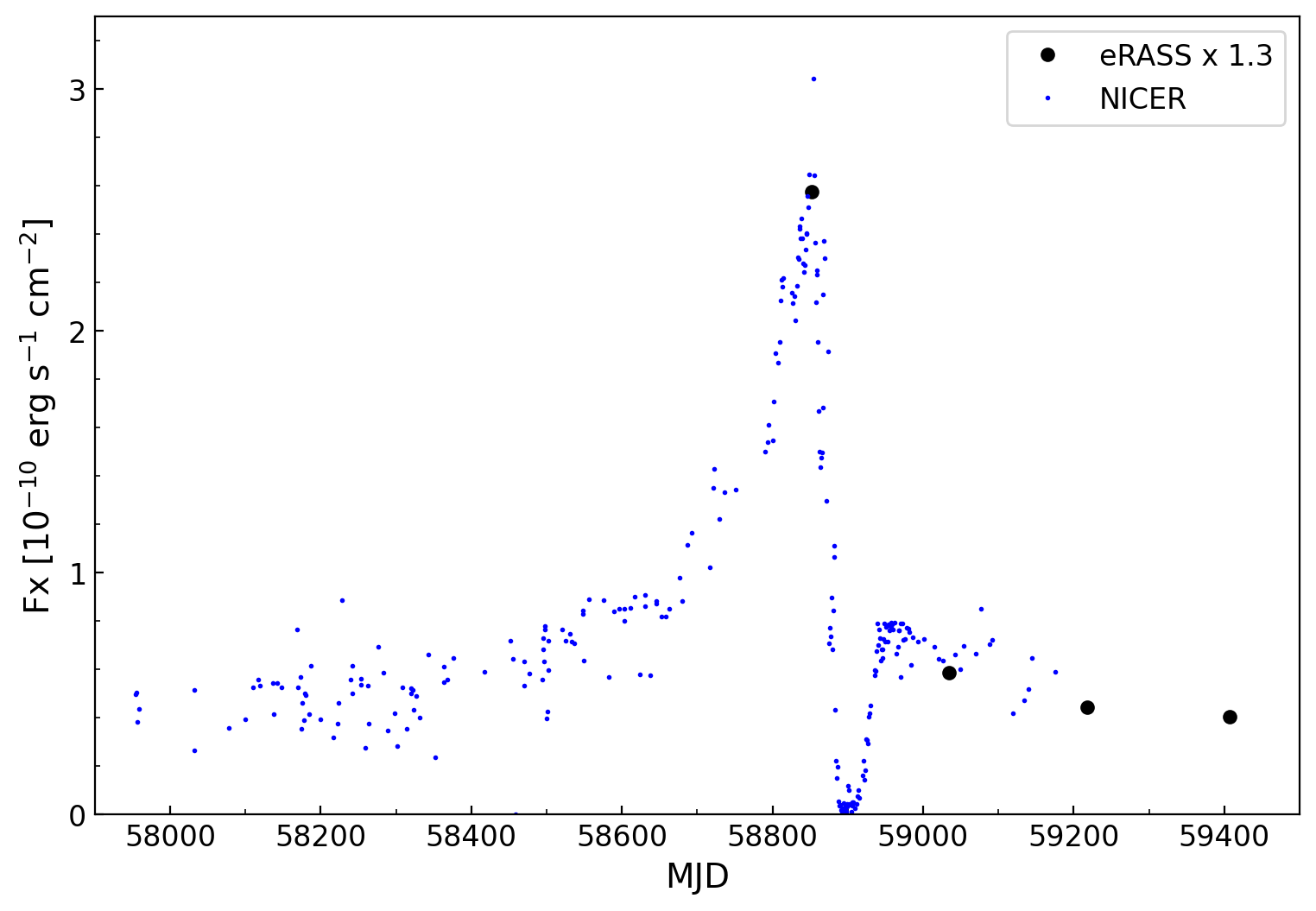}
\end{center}
\caption{\label{erassvsnicer} X-ray lightcurve of $\eta$ Car showing the 2.0\,--\,10.0 keV flux as observed with NICER and eROSITA. Note that the eRASS flux was scaled up by 30\,\% to broadly match the NICER results taken from \cite{2022ApJ...933..136E}.}
\end{figure}

The object $\eta$~Carinae has been monitored in X-rays over several decades. The flux variability and spectral changes observed during the eRASS are similar to those observed over previous orbits \citep{2017ApJ...838...45C} and overall agree with findings from NICER observations around the 2020 periastron passage \citep{2022ApJ...933..136E}. However, the eRASS data give about 30\,\% lower fluxes in the 2.0\,--\,10.0~ keV band when compared to the NICER values. Although the data were not taken strictly simultaneously, this indicates cross-calibration issues potentially caused by effective area uncertainties at energies above $\approx 2$~keV for the eROSITA survey data. A comparison of the derived fluxes from the two instruments is shown in Fig.~\ref{erassvsnicer}.
Given the errors, the eRASS data are well described by a two-temperature model with $kT1 = 0.8 - 1.0$ keV and $kT2 = 4.0 - 4.5$~keV. Notably, the eRASS2 data at phase 0.068 are better described by a temperature of roughly 3~keV and similarly the NICER spectral models give 3.0\,--\,3.6~keV for data taken within a few days. However, it is also during a phase of quite strong absorption, while $kT$ and $N_{\rm H}$ are interdependent parameters in spectral modelling.

The eROSITA spectrum of $\eta$~Car is predominantly of thermal origin as shown in our model fits and evidenced by the strong \ion{Fe}{XXV} 6.7~keV emission line. Surely, a fainter non-thermal contribution can always be `hidden', especially in the spectra with a moderate signal-to-noise ratio, as its basic spectral shape for reasonable values of the photon index is similar to those of a hot thermal plasma, but it is not required to adequately describe our data nor favored in the fits. On the other hand,  non-thermal emission from $\eta$~Car is detected at hard X-rays and soft Gamma-ray energies \citep{2018NatAs...2..731H}. However, the expected contribution is only at a percent level in the eROSITA energy range, consistent with our results and conclusions from the spectral analysis.

\subsection{Wolf-Rayet stars}

The CNC harbours three additional X-ray bright Wolf-Rayet stars, WR~22, WR~24, and WR~25, which are also eRASS sources. We discuss the data and summarise the spectral modelling results in Tab.~\ref{stars_s}.

\paragraph{WR 25 (HD~93162):} This star is a massive colliding wind binary of O2.5If*/WN6+O type. It has component masses of about 75+30~$M_{\odot}$, an orbital period of 208~d period and has been studied in pointed X-ray observations, see e.g. \cite{2003A&A...402..653R, 2006A&A...460..777G, 2006A&A...445.1093P}. In our survey data, WR~25 is the X-ray brightest among the Wolf-Rayet stars and after $\eta$~Car the second brightest X-ray source in the studied Carina field. During eRASS, WR~25 was observed at orbital phases of 0.90, 0.77, 0.66, and 0.57, and its X-ray brightness varies by a factor of about two as shown in Fig.~\ref{stars_lcs}. We measure an observed $F_{\rm X} = 3.3 - 6.9 \times 10^{-12}$~erg\,cm~$^{-2}$\,s~$^{-1}$, with the highest flux detected in the pre-periastron phase. However, changes in the spectral properties are minor and beside a re-normalisation, a uniform spectral model is suitable to describe the data. We find an average $\log L_{\rm X} = 33.6$ and plasma components of comparable strength at temperatures of $kT \approx 0.8$ and 2.0~keV.

\paragraph{WR~22 (HD~92740):} The Wolf-Rayet star WR~22 is a WN7h+09III-V system with 70+25~$M_{\odot}$ and $P_{\rm orb} =80$~d. A discussion of the WR~22 system and a phase-resolved X-ray study with {\it XMM-Newton} is presented in \cite{2009A&A...508..805G}. Over the eRASS, X-ray flux variations due to the orbital motion of about a factor of two are seen, see Fig.~\ref{stars_lcs}. The phase dependence (eRASS1-4 phases are 0.59, 0.86, 0.15, 0.50) of the observed rate is comparable to the one observed with {\it XMM-Newton}. The spectra of WR~22 show a dominant 0.6\,--\,0.7~keV plasma component. In addition, we find indications for contributions from hotter plasma with $kT \gtrsim 2.0$~keV as in WR~25, but this component is poorly constrained by our data.

\paragraph{WR~24 (HD~93131):} WR~24 is a Wolf-Rayet star of spectral type WN6ha-w and the X-ray faintest among the studied WR stars. Its plasma temperature is comparable to those of WR~22 and similarly indications for a hotter plasma component are present. These properties are quite typical for a colliding wind binary, however so far no companion has been identified. While strong intrinsic absorption is seen in all WR stars, it is highest in WR~24.

\subsection{X-ray bright O stars}

Other X-ray bright stellar sources are detected among the O-type stars, which are typically the more massive members with very strong stellar winds in a binary or a multiple system. An overview of the population of massive stars and a description of individual sources is given in \citet{2011ApJS..194....5G}, from where we adopt the stellar parameters if not denoted otherwise. We model and discuss the eRASS data from the most outstanding X-ray bright stellar sources in the following.

\paragraph{HD~93129 (Trumpler 14-1):} The O-type star HD~93129 is among the earliest-type stars in the Milky Way and a bright X-ray source. It was classified as an binary of the type O2If*+O3.5V((f))5; however, HD~93129~A was discovered to be a binary itself consisting of an O2 supergiant and an O3.5 star, making HD~93129 at least a triple system. There is another massive binary HD~93128 (Trumpler 14-2) nearby, which is a spectroscopic binary of the spectral type O3.5V((fc)). Both stars are separated by 24~arcsec and not fully resolved in the eRASS data, with HD~93129 being the dominant X-ray source by a factor of about four. Strong X-ray emission from hot plasma at $kT$ of a few keV is detected from this source.

\paragraph{HD~93160 and HD~93161:} These are O-type multiples of the spectral type O7III((f)) SB and O8V+O9V+O6.5(f) located in Trumpler~14. They are separated by 15~arcsec and blended in eRASS, the flux ratio between the two sources is about one with HD~93161 being moderately brighter by 20\%. 

\paragraph{HD~93205 (V560 Car)} is an O-type binary of the spectral type O3.5V((f))+O8V. Its X-ray spectrum is among the softest of the studied sources and we find that plasma at temperatures up to 1~keV dominates its X-ray emission.

\paragraph{HD~93250} is an O4III(fc) star with suspected binarity, but a secondary was not unambiguously identified so far. The eRASS data confirms it as a bright X-ray source at $L_{\rm X} = 10^{33}$ erg s$^{-1}$ with a strong hot plasma component at $kT$ of a few keV.

\paragraph{HD~93403} is an O5.5III(fc)+O7V eccentric binary. It is a bright X-ray source with $L_{\rm X} = 1.3 \times 10^{33}$ erg s$^{-1}$ and has a hard X-ray spectrum that shows strong emission from hot plasma at $kT$ of a few keV.

\paragraph{HD~93206 (QZ Car)} located in Collinder~228 is a O9.7I+O8III double binary, i.e., at least a quadruple system. Bright X-ray emission with a hot plasma component is observed from this massive O-type multiple system.

\paragraph{Coll~228-103} is an OB star in the southern cluster Collinder~228, which is part of an OB mini-cluster consisting of Coll~228 101-103 and HD~305453.

\paragraph{} In the studied O-type stars the X-ray emission typically exhibits a strong and often dominant plasma components at temperatures of about 0.6\,--\,0.8~keV. In addition, in several targets, significant contributions from hotter plasma at a few keV is present, most prominently in HD~93129, HD~93206, HD~93250, and HD~93403, which all have $kT_{\rm av} \gtrsim 1$~keV. These high temperatures are specifically found in colliding wind binaries or other multiple systems. In contrast, a few targets have predominantly cooler plasma in addition to the 0.6\,--\,0.8~keV component (HD~93205, Coll~228-103) and exhibit $kT_{\rm av} \lesssim 0.5$~keV in our models. Comparing the four eRASS visits for the O-type stars, we find only mild variability within about 20\,\% in the observed X-ray fluxes. All spectral modelling results of the studied sources are summarised in Tab.~\ref{stars_s}.

\begin{table*}[t]
\caption{\label{stars_s}Spectral fit results for eRASS:4 data. $F_{\rm X}$ is the observed 0.2\,--\,5.0~keV flux, $L_{\rm X}$ is the absorption corrected 0.2\,--\,5.0~keV X-ray luminosity.}
\begin{center}
\begin{tabular}{lrrrrrrr}\hline\hline\\[-3.mm]
name  & $N_{\rm H}$ &  $kT_{1}$ & $\log EM_{1}$ & $kT_{2}$ & $\log EM_{2}$ & $F_{\rm X}$ & $\log L_{\rm X}$ \\
 & 10$^{22}$cm~$^{-2}$ &  keV & [cm$^{-3}$] &  keV & [cm$^{-3}$]  &  erg\,cm$^{-2}$\,s$^{-1}$ & [erg\,s$^{-1}$] \\\hline\\[-3mm]
WR 22 &   0.73\,$^{ +0.18}_{ -0.19}$ &   0.64\,$^{ +0.11}_{ -0.13}$ &  55.45\,$^{ +0.21}_{ -0.22}$ & -- & -- & 2.03e-13 &  32.27 \\ 
WR 24 &   1.38\,$^{ +0.34}_{ -0.24}$ &   0.68\,$^{ +0.16}_{ -0.17}$ &  55.72\,$^{ +0.26}_{ -0.21}$ & -- & -- & 1.77e-13 &  32.14 \\ 
WR 25 &   0.78\,$^{ +0.08}_{ -0.10}$ &   0.76\,$^{ +0.06}_{ -0.05}$ &  56.41\,$^{ +0.11}_{ -0.14}$ &   1.88\,$^{ +0.36}_{ -0.22}$ &  56.46\,$^{ +0.05}_{ -0.06}$ & 4.68e-12 &  33.56 \\ 
HD 93129/8 &   0.23\,$^{ +0.07}_{ -0.03}$ &   0.85\,$^{ +0.08}_{ -0.07}$ &  55.44\,$^{ +0.12}_{ -0.12}$ &   3.79\,$^{ +2.75}_{ -1.19}$ &  55.95\,$^{ +0.07}_{ -0.08}$ & 1.93e-12 &  33.25 \\ 
HD 93160/1 &   0.52\,$^{ +0.49}_{ -0.30}$ &   0.67\,$^{ +0.11}_{ -0.15}$ &  55.32\,$^{ +0.08}_{ -0.08}$ & -- & -- & 2.26e-13 &  32.34 \\ 
HD 93205 &   0.20\,$^{ +0.10}_{ +0.00}$ &   0.13\,$^{ +0.30}_{ -0.08}$ &  55.13\,$^{ +2.16}_{ -0.70}$ &   0.74\,$^{ +0.20}_{ -0.08}$ &  55.12\,$^{ +0.06}_{ -0.27}$ & 3.32e-13 &  32.78 \\ 
HD 93206 &   0.20\,$^{ +0.16}_{ -0.00}$ &   0.71\,$^{ +0.05}_{ -0.11}$ &  55.07\,$^{ +0.22}_{ -0.12}$ &   1.52\,$^{ +0.90}_{ -0.30}$ &  55.23\,$^{ +0.12}_{ -0.15}$ & 5.34e-13 &  32.77 \\ 
HD 93250 &   0.20\,$^{ +0.08}_{ -0.00}$ &   0.80\,$^{ +0.07}_{ -0.07}$ &  55.19\,$^{ +0.13}_{ -0.08}$ &   4.48\,$^{+11.7}_{ -1.85}$ &  55.67\,$^{ +0.08}_{ -0.09}$ & 1.10e-12 &  33.03 \\ 
HD 93403 &   0.27\,$^{ +0.13}_{ -0.07}$ &   0.80\,$^{ +0.05}_{ -0.05}$ &  55.49\,$^{ +0.04}_{ -0.05}$ &   2.02\,$^{ +0.99}_{ -0.37}$ &  55.68\,$^{ +0.06}_{ -0.06}$ & 1.27e-12 &  33.08 \\ 
Coll 228-103 &  0.20\,$^{ +0.00}_{ -0.00}$ &  0.27\,$^{ +0.25}_{ -0.13}$ &  54.72\,$^{ +0.31}_{ -0.35}$ &   1.14\,$^{ +0.33}_{ -0.23}$ &  54.73\,$^{ +0.15}_{ -0.20}$ & 1.57e-13 &  32.34 \\\hline
\end{tabular}
\end{center}
\tablefoot{$N_{\rm H}$ is the total column density, with the minimum value set to the mean ISM value. A two-temperature fit was used when it improved the fit significantly.}
\end{table*}

%-----------------------------------------------------------------
\section{Discussion}

The CNC-Gum 31 region is a very bright \hii\ region and a very active 
star-forming complex, but not as extreme as 30 Doradus (Dor) in the Large 
Magellanic Cloud \citep{1998ApJ...493..180M} or the Arches cluster near the 
Galactic Centre \citep{2004ApJ...611L.105N}.
However, the close location of the CNC-Gum 31 complex  and its properties make 
it an excellent target for the study of high-mass star-forming regions and
therefore also for the understanding of the star formation in the early Universe.

\subsection{Multi-wavelength comparison}\label{disc:mwl}

In a new \hi\ study at  the Australia Telescope Compact Array, 
complex filamentary structures were found in the distribution of the
cold neutral gas over a wide range of velocities
\citep{2017MNRAS.472.1685R}.
Several `bubbles' were detected in the CNC, which have
obviously been produced by the impact of the massive stars.

High-resolution observations of the $^{12}$CO(1-0) and $^{13}$CO(1-0) molecular lines 
in the Carina nebula and the Gum 31 region obtained with the 22~m Mopra 
telescope as part of The Mopra Southern Galactic Plane CO Survey was
presented by \citet{2016MNRAS.456.2406R}.
From the sample of the massive clumps observed with Mopra, two regions with
very different properties have been observed at the Atacama Large 
Millimeter/Submillimeter Array (ALMA) by \citet{2020ApJ...891..113R}.
One region is located at the centre of the nebula and has most likely 
experienced some feedback from the massive star clusters, 
while the other region was selected further south and is believed to be 
less disturbed by the massive stars.
It was shown that the centre of the nebula is forming  fewer cores than in
the south, however, the cores are much more massive. These differences
suggest that the stellar feedback has an effect on the formation of clumps
which evolve into star-forming cores. 
In addition, \citet{2021PASJ...73S.201F} have analysed data taken with the
Mopra and NANTEN2 telescopes as well as with ALMA and have detected
signatures of cloud-cloud collisions in the velocities and the morphologies at 
different velocities of the clouds. They suggest that the collisions between
clouds have triggered massive star formation in CNC.

Moreover, the radio continuum map in the 1 -- 3 GHz range 
\citep{2021ApJ...909...93R}
shows a complex distribution of emission across the nebula consisting of 
filaments, shells, and fronts on different size scales.
Ionization fronts are found at distances of 80 pc from the stellar clusters in 
the centre of the CNC to the north and the south.
It has been estimated that approximately 15\% of the ionising flux has escaped 
from the nebula into the diffuse Galactic interstellar medium. 

From the comparison of the radio continuum emission to the diffuse X-ray 
emission we can conclude that
the cavities seen in the \hi\ map in the north and the south have most likely 
been formed by the influence of massive stars in the star clusters at the 
southern end of the northern cloud, also confirmed by the bright diffuse X-ray 
emission in the Central region C2.
As pointed out by \citet{2021ApJ...909...93R},
the radio continuum and IR maps show plumes at the edges of the \hii\ region,
also around the northern tip of the X-ray hook. These plumes are indicative
of cavities filled with over-pressurised hot gas, which broke out and is 
expanding out of the molecular cloud into the Galactic plane.
The diffuse X-ray emission in the OS2 region apparently is due to the expanding
hot plasma in the southern cavity.

\subsection{Origin of the X-ray emission}\label{disc:origin}

The results of the spectral analysis indicate that the emission is mainly caused by thermal plasma in all regions. The element abundances are consistent with solar values, which indicates that the origin of the hot plasma is shocked ISM. \citet{2011ApJS..194...15T} found enhanced Fe abundance in their `central arc' region, which corresponds to our Hook region H1. In  young stellar clusters like those in the Carina nebula it is likely that there has already been a supernova explosion and that an SNR contributes to the X-ray emission. The combination of stellar winds and SNRs will create a superbubble around the stellar clusters. Therefore, in the context of the multi-wavelength study, we compare the observed emission with superbubble simulations \citep{Krausea13a,Krausea14a}.

The bright X-ray hook region is most likely caused by a shock in a mixing region at the edge of a superbubble driven mainly by $\eta$ Car. The time variability of the stellar driving power of superbubbles leads to various instabilities of superbubble shells, which tend to produce more extended turbulent mixing regions compared to the gradual change in density and temperature from the hot bubble interior to the dense shell, which would be expected from thermal conduction alone. Dissipation in such mixing regions dominates the radiative losses in superbubbles \citep[compare][]{Fierlea16}. Contrary to early 1D models with steady source power \citep[e.g.,][]{CM90} the losses can be strong enough,
hence the internal energy density in the superbubble low enough, such that a significant increase in stellar source power can drive a shock wave through the entire superbubble. In the simulations of \citet{Krausea14a}, supernova shock waves encountered the optimal conditions in such mixing regions towards the inner edge of supershells to produce an order of magnitude enhancement in soft X-rays.
In the present case, it seems likely that $\eta$ Car, currently in the LBV phase, drives such a shock. An LBV phase is expected to last about $10^5$~years \citep{Grohea14} with strong mass loss
and powerful eruptive, possibly explosive winds \citep{Gormea22,Vink22}. Massive star winds have typical velocities of the order of 1000~km~s$^{-1}$. The distance of $\sim$10 pc between $\eta$ Car and the X-ray hook are hence traversed in about $10^4$~yr.
It is therefore likely that the X-ray enhancement in the Hook region is caused by $\eta$ Car's current LBV phase.

The Central regions around $\eta$ Car are
best explained by the classical inner shock of its wind \citep{Weavea77}. Due to strong adiabatic expansion, the temperature of the gas before the shock is low, hence the shock's Mach number is high. It is therefore highly likely to find diffusive shock acceleration at such shocks \citep{VY14}. Consequently, we find emission from non-thermal electrons in the regions around the massive stars only, including $\eta$ Car.

The radio continuum image \citep{2021ApJ...909...93R} 
 suggests a superbubble around $\eta$ Car, which is 
limited towards the south-east by the Hook region and extends significantly further towards the north-west. The latter is confirmed by the HI map in Fig.~\ref{images}. Our Outer region O1 extends across a significant part of this superbubble, as defined by the radio continuum image, and is co-spatial with the radio continuum as well as the $R$-band emission. Both, radio continuum and $R$-band image show photoionised gas, which is found towards the inner edge of the superbubble shells.

The situation is different for our Outer region O2. There, the X-ray-bright region as well as the photoionised gas region is much narrower in the direction of the major axis of the corresponding superbubble (i.e., the one south of the Hook region). The X-rays here are anti-correlated with the photoionised gas, with X-rays being further inside the superbubble.

Both our outer regions can be explained by hot gas being pushed against the shell of their respective superbubbles.
The compression, and possibly some mixing, enhances the emission measure, and the regions are thus more prominent in X-rays. Southwards the superbubble extends close to the plane of the sky. It is the comparatively denser part of the shell that is photoionised. Hence, the X-rays appear ``inside'' the shell that is delineated by the photoionised gas, similar to the case of the Orion-Eridanus superbubble \citep{Krausea14a}. The region O1 is seen almost ``from the top'' (or bottom), i.e., the superbubble extends at a significant inclination to the plane of the sky. Hence, we see the X-ray-bright and the radio continuum-bright regions
on top of each other.

The fact that we see both Outer regions O1 and O2 at similar X-ray brightness at the same time is probably accidental. As shown in \citet{Krausea14a}, the temporal variability of the stellar sources inside superbubbles leads to gas sloshing, i.e., the hot gas moves around in superbubbles. 
The expected timescale of this variability is about $10^5$~yr.
Wherever the hot gas hits the dense shell, it becomes brighter due to the adiabatic compression. This interpretation is consistent with the many arcs in the radio continuum image that are not associated with X-ray-bright regions. It is also possible that the Outer region O1 is indicative of a temporal compression in the interior of the superbubble related to the gas sloshing.

The different bubbles around the individual massive star groups in the CNC are highly likely to have very different pressures. This follows from the strong dependency of the wind power on mass and evolutionary state of a star. The $\eta$~Car bubble is clearly the highest-pressure system, which is not only evident from it having the most powerful stellar source, $\eta$~Car, but also from the shock it drives, visible as the Hook region.

Due to the overpressure in the $\eta$~Car bubble, a hot wind is driven through the molecular clouds in the centre of the system. The wind has swept away the more tenuous parts of the molecular cloud with which it is mixed southwards in our Outskirts regions. The mixing decreases the X-ray temperature and increases the emission measure, as observed. It is also interesting to compare
the HI map (Fig.~\ref{images}) to the molecular and radio continuum data 
\citep[e.g., ][their Fig. ~9]{2021ApJ...909...93R}.
While the northwest superbubble and the southern superbubble seem connected in the HI map, molecular
as well as ionised gas seems to be left in the interface
between the two superbubbles. This could mean that the intermediate-density gas (HI) is blown away
and mixed into the X-ray gas, whereas the denser,
molecular parts remain. Some of the dense gas is photoionised and keeps being ablated.

The denser gas in the molecular clouds, which is not swept away, is likely to be surrounded by the hot X-ray gas and compressed by its high pressure
\citep[compare][]{Krausea18b}. This may well be the mechanism that formed the massive cores detected by ALMA in the central part of the nebula (compare above). This
process has been suggested to explain the sequential star formation in the Scorpius-Centaurus superbubble by \citet{Krausea18b}. Their simulations directly support the combination of sweeping away of more tenuous gas combined with the formation of dense gas cores. The plumes identified in the radio continuum image are also consistent with this picture, probably representing somewhat denser gas phases that are pushed out of the molecular region by the hot wind.

\subsection{Thermal properties of the superbubble}

If the high-mass stars in the central region blow strong stellar winds and encounter an obstacle in the X-ray hook, these regions should have a higher pressure than the surroundings. We use the results from the X-ray spectral analysis to estimate the density and pressure in these regions. 

We assume that in the central region, the stars blow a bubble with a size similar to the Central regions C1 and C2. We approximate the bubble as a sphere with a diameter of 0.37\degr. The Hook region (H1 and H2) is most likely a shell-like structure around the central region, which we see from the side. We assume that it can be modelled as a part of a spherical shell with an outer diameter of 1\degr\ and an inner diameter of 0.65\degr. We will also study the Outskirt region OS2, which is plasma flowing into the northern part of the southern cavity. We model the structure as a half ellipsoid with a major axis $a$ of 0.64\degr\ and a minor axis $b$ of 0.5\degr. For the axis parallel to our line of sight $c$, we assume the same value as for the minor axis $b$. 

Based on these assumptions, we calculate the following volumes for the X-ray emission:
for the Central regions (C1 and C2) $V_{\mathrm c}$ = 4.2 $\times 10^{58}$ cm$^3$,
for Hook regions (H1 and H2) $V_{\mathrm h}$ = 2.9 $\times 10^{59}$ cm$^3$, and 
for the region OS2 $V_\mathrm{os}$ = 5.8 $\times 10^{59}$ cm$^3$.

To calculate the density, we use the normalisation of the spectral models for thermal plasma:
\begin{equation}
    norm_i = \frac{10^{-14}}{4 \pi D^2} \int n_e n_{\mathrm H} dV = \frac{10^{-14}}{4 \pi D^2} 1.2 n_{\mathrm H}^2 V_i,
\end{equation}
with $D$ being the distance to the CNC and assuming homogeneous density in each of the regions and $n_e = 1.2 n_{\mathrm H}$.
The index $i =$ c, h, and os denotes the parameters for the Central regions (combined), Hook regions (combined), or the OS2 region.
Therefore, for the hydrogen density and the pressure of the plasma, we get
\begin{equation}\label{eq:dens}
    n_{{\mathrm H}, i} = \sqrt{\frac{4 \pi D^2 norm_{i}}{1.2 x 10^{-14} V_{i}}}
\end{equation}
and
\begin{equation}\label{eq:press}
    P_{i}/k = (n_{e, i} + n_{{\mathrm H}, i} + n_{\mathrm{others}, i}) T = (1.2 + 1.1) n_{{\mathrm H}, i} T_{i}.
\end{equation}
Using Eq.\,\ref{eq:dens} and \ref{eq:press} we get 
for the Central regions % use the hot component in Central 2
$n_\mathrm{H, c}$ = 4.0 $\times 10^{-2}$ cm$^{-3}$ and $P_{\mathrm c}/k$ = 9.1 $\times 10^{5}$ cm$^{-3}$ K,
for the Hook regions % use the hot component in Hook 1 
$n_\mathrm{H, h}$ = 5.4 $\times 10^{-2}$ cm$^{-3}$ and $P_{\mathrm h}/k$ = 9.4 $\times 10^{5}$ cm$^{-3}$ K, and
for the OS2 region % use the hot component in Hook 1 
$n_\mathrm{H, os}$ = 2.3 $\times 10^{-2}$ cm$^{-3}$ and $P_\mathrm{os}/k$ = 5.1 $\times 10^{5}$ cm$^{-3}$ K.
The uncertainties of the temperature and the normalisation are about 5\% and 10\%, respectively, while our volume estimate is most likely uncertain by 20 -- 30 \%. Therefore, the derived values have an uncertainty of $\sim$50\%. The numbers still indicate that the density and the pressure are higher in the Central and Hook regions, while the lower pressure in the southern cavity indicates outflowing gas. The pressure in the CNC is one to two orders of magnitude higher than the average pressure in the ISM in the Milky Way \cite[][and references therein]{2021ApJ...916...17J}.

\subsection{Cosmic ray pressure}\label{ssec:crpress}

In the shocks of the strong stellar winds of massive stars, electrons and protons are accelerated to relativistic energies and become cosmic rays. In young stellar clusters like the Carina Nebula, there also regions in which stellar winds collide and which are thus very efficient acceleration regions \citep[see, e.g.,][and references therein]{2023arXiv230903746S}.
The harder emission in the spectrum in the Central regions
can be either explained by synchrotron radiation or inverse
Compton scattering of the accelerated particles. 

For the synchrotron hypothesis, we need to assume a magnetic field strength. As the thermal pressure is about two orders of magnitude higher than the Galactic average, we might expect
a similarly enhanced magnetic pressure. Since the magnetic field in the Milky Way is generally of the order of $\mu$G
\citep{SR10,XH19,Dickea22}, we may expect a magnetic field in the Carina superbubble system of several 10~$\mu$G.
The synchrotron emissivity
for a power-law distribution of electrons is given by \citep{Longair2011}:
\begin{equation}
J(\nu) = A\frac{\sqrt{3\pi}e^3B}{16\pi^2\epsilon_0cm_e(q-1)} \kappa\ff{2\pi \nu m_e^3 c^4}{3eB}^{-\frac{q-1}{2}}
\end{equation}
with
\begin{equation}
A=\frac{
\Gf\nbr{\frac{q}{4}+\frac{19}{12}} 
\Gf\nbr{\frac{q}{4}-\frac{1}{12}}
\Gf\nbr{\frac{q}{4}+\frac{5}{4}}
}{\Gf\nbr{\frac{q}{4}+\frac{7}{12}}}\, .
\end{equation}
Here, 
$m_e$ and $e$ are, respectively, electron mass and charge, $c$
is the speed of light and $\epsilon_0$ the vacuum permittivity. The Gamma function is denoted by $\Gf$
and the electron energy distribution is $n(E)=\kappa E^{-q}$. The electron distribution power-law index $q$ is related to the photon index $\Gamma$ by $q=2\Gamma-1$.
For an assumed magnetic field strength $B$,
the characteristic frequency for synchrotron emission
is given by 
\begin{equation}
\nu_\mathrm{c}=7\times10^{17}\,\mathrm{Hz} \ff{E}{30\,\mathrm{TeV}}^2 \ff{B}{\mathrm{nT}} \, .
\end{equation}
The non-thermal luminosity observed
in the X-ray band is given by the integral:
$\int_{\nu_1}^{\nu_2} J(\nu)\,\mathrm{d}\nu\,$ ,
and we take $\nu_1 = 4.5\,$keV$/h$ and $\nu_2 = 12\,$keV$/h$ to be clearly dominated by the 
non-thermal part of the spectrum.
This procedure allows us to infer $\kappa$ for
an assumed magnetic field by equating the predicted luminosity to the non-thermal luminosity we observe in regions Central~1 and~2, $L_\mathrm{obs,nt,4.5-12\,keV} = 5.59\times10^{33}\,$erg$\,$s$^{-1}$, which we use to infer
the relativistic electron pressure by integrating
over the energy distribution, 
\begin{equation}
    P_\mathrm{cr,e} = (\gamma_\mathrm{ad}-1) \int_{E_1}^{E_2}
    E \, n(E)\,\mathrm{d}E\, ,
\end{equation}
where the adiabatic index is $\gamma_\mathrm{ad} = 4/3$ for 
relativistic particles.
We follow \citet{delPalea22} in adopting a leptonic 
contribution to the overall cosmic ray pressure of 
$2\,$\%.

\begin{figure}
\begin{center}
\includegraphics[width=0.5\textwidth]{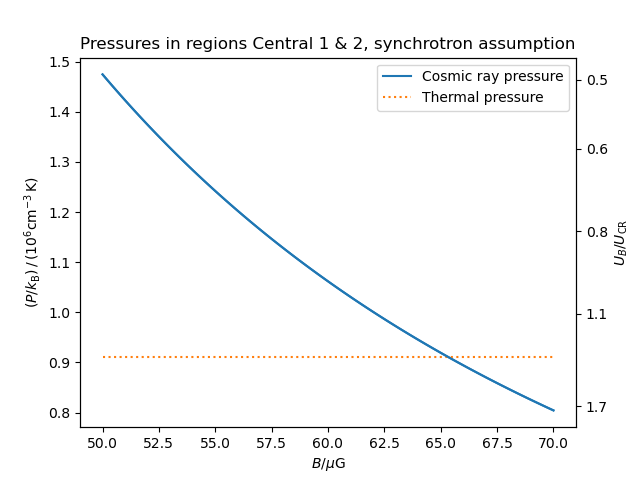}
\end{center}
\caption{\label{fig:synch} Possible cosmic ray pressure in regions C1 and C2 under the assumption that the measured non-thermal X-ray radiation is synchrotron emission. The blue line shows the total cosmic ray pressure (hadrons $+$ leptons, left vertical axis) and respectively the ratio between magnetic and total cosmic ray pressure (right vertical axis) over magnetic field strength. The magnetic field values used are not observationally known, but are reasonable values given other observables of the region. The orange line shows the thermal pressure of the same region for comparison (left axis). See Sect.~\ref{ssec:crpress} for 
a discussion of this plot.
}
\end{figure}

The cosmic ray pressure as well as the ratio between the magnetic 
and the cosmic ray pressure are shown over the assumed magnetic field strength in Fig.\,\ref{fig:synch}. The thermal pressure is also 
shown for comparison. Under the synchrotron hypothesis, our measurements allow for two possible solutions: if the magnetic field strength is low, the region must be dominated by the cosmic ray pressure. If the magnetic field is high, the magnetic pressure will dominate. The measurements exclude the possibility that the thermal pressure dominates. A solution with all three pressure components in approximate equipartition exists for $B\approx60\,\mu$G. We note that for these assumptions, we predict a non-thermal radio flux at 2~GHz of 7~Jy, well consistent with the thermally dominated radio continuum emission of 2~kJy observed by \citep{2021ApJ...909...93R}. There are generally very few detections of non-thermal radio components in superbubbles, and it is possible that special geometries or circumstances may be required to distinguish the non-thermal contributions \citep[compare][]{Heesea15}.

Strong magnetic field amplification is 
expected for shocks that produce the 
high-energy leptons (up to about 30~TeV) required in the synchrotron hypothesis \citep{Schurea12,Vink12}. The physics 
could be similar to the situation in the
30~Dor superbubble \citep{Kavanea19}, where TeV emission of the high-energy lepton population may be observed, though in the present case the accelerating shock would be the terminal wind shocks of the massive stars in the Central regions. Gamma-rays  up to at least 400~GeV are also observed from the direction of $\eta$~Car
\citep{2010ApJ...723..649A,2021A&A...654A..44M,HESS20,2023arXiv230903746S}. 
Given the extent of the point spread function in the HESS detection,
some of this emission might be related to
any inverse Compton component of the hypothetical TeV electrons, though the correlation of the GeV emission with $\eta$~Car's orbital period observed with Fermi \citep{2021A&A...654A..44M} and the detection of the colliding wind binary in hard X-rays with NuSTAR associate the bulk of this high-energy emission directly with $\eta$~Car \citep{2018NatAs...2..731H}. The GeV $\gamma$-ray emission observed with \fermi\ \citep{2010ApJ...723..649A,2021A&A...654A..44M} could also have a contribution from leptonic emission \citep{HESS20}, but seems at least
partly to be due to pion decay caused by the interaction of a hadronic
cosmic ray population with the dense gas in the region \citep{Gea22}.

The possible non-thermal X-ray spectrum we detect could  be caused by inverse Compton upscattering of starlight or photons from the cosmic microwave background (CMB) by relativistic electrons with much lower Lorentz factors.
The frequency shift is approximately given by 
\citep{Longair2011}:
\begin{equation}
    \frac{E_\mathrm{IC}}{5\,\mathrm{keV}}
    = \ff{\gamma}{30}^2 \ff{400\,\mathrm{nm}}{\lambda}
    = \ff{\gamma}{2000}^2 \ff{\nu}{160\,\mathrm{GHz}}\, .
\end{equation}
Hence, upscattering from optical / UV (wavelength $\lambda \approx 400\,$nm) to X-rays requires electrons with Lorentz factors of  
$\gamma\approx30$. 
From the direct measurements of the local cosmic ray electron population at MeV energies,
we know that the energy spectrum becomes very flat, $q\approx 1.4$ \citep[after accounting for the solar modulation,][]{Orlando18,Vittea19,KH21}. If the situation was similar in the CNC,
one would expect a non-thermal photon index of around 1.2 rather than 1.8, which we observe.
We estimate the cosmic ray pressure for this case in the following way: for a radiation energy density $U_\mathrm{rad}$ the inverse Compton luminosity is given by \citep{Longair2011}:
\begin{equation}
    L_\mathrm{IC} = \frac{4}{3}\gamma^2\sigma_\mathrm{T}cU_\mathrm{rad} N_\mathrm{e}\,
\end{equation}
with Thompson cross section $\sigma_\mathrm{T}$ and the number of upscattering electrons $N_\mathrm{e}$. The latter can then be obtained by equating $L_\mathrm{IC}$ to the observed luminosity. The total cosmic ray pressure for the same assumptions as above is then given by:
\begin{equation}
    P_\mathrm{CR} =
    (\gamma_\mathrm{ad}-1) 
    \frac{51 N_\mathrm{e} \gamma m_\mathrm{e}c^2}{V}\,
\end{equation}
\citet{Smith06} gives a luminosity in ionising photons of
$Q=10^{51}\,$s$^{-1}$, corresponding to a photon density of a few photons per cm$^{3}$ or $U_\mathrm{rad}\approx 10^{-11}\,$erg$\,$cm$^{-3}$. This would yield a cosmic ray pressure of $P_\mathrm{CR}/k\approx 10^9\,$cm$^{-3}\,$K, much higher than what we would expect in this region (compare above).

The CMB peaks at a frequency of $\nu=160\,$GHz and has an energy density 
of $U_\mathrm{rad}=4\times10^{13}\,$erg$\,$cm$^{-3}$.
A similar analysis as before yields an estimate of the total cosmic ray pressure of $P_\mathrm{CR}/k\approx 10^9\,$cm$^{-3}\,$K, similar to the above estimate and again much higher than the thermal pressure, which we derive above. 

In conclusion, if the non-thermal spectrum in the Central regions was due to inverse Compton scattering, the pressure in the region would be strongly dominated by cosmic rays, even if only the leptonic component was taken into account. These considerations seem to favour the synchrotron hypothesis. More sensitive TeV observations might be able to confirm this or otherwise.

As a further consistency check, we can work out 
the radius of the termination (inner) shock of the
combined wind from the $\eta$~Car binary. We adopt 
the values collected in \citet{Grohea12}
(see  Table~\ref{stars_w}).
With $\rho$ being the wind density and $v$ the velocity of the wind, the wind ram pressure equals the ambient pressure $p= \rho v^2$ at the termination shock. We  take the thermal pressure in the central region as proxy for the ambient pressure. With this assumption, we find for the radius of the termination shock:

\begin{equation}
    r_\mathrm{ts} = \nbr{\frac{L\dot M}{8\pi^2 p^2}}^{1/4}
    = 15\,\mathrm{pc}\,\ff{L}{L_\mathrm{c}}^{1/4}
    \ff{\dot M}{\dot M_\mathrm{c}}^{1/4}
    \ff{p}{P_\mathrm{c}}^{-1/2}.
\end{equation}

Given the evidence for the additional cosmic 
ray pressure, we expect a termination shock somewhat closer to $\eta$~Car than 15~pc,
consistent with our assumption above that it is
this shock that accelerates the non-thermal particles to TeV energies in the Central regions, and the synchrotron
hypothesis above. Given this magnitude of the termination shock radius, it is possible that the shock has traversed the entire distance between $\eta$~Car and the denser gas in the Hook region.

\begin{table}[t]
\caption{\label{stars_w}Properties of some massive-star winds in the CNC}
\begin{center}
\begin{tabular}{lcccc}\hline\hline\\[-3.mm]
name  & $\dot M/10^{-5}$ &  $v_\infty$ & $L_\mathrm{kin}/10^{37}$ & Refs. \\
 &  $M_\odot\,$yr$^{-1}$ &  km$\,$s$^{-1}$ & erg$\,$s$^{-1}$ &     \\\hline\\
$\eta$~Car~A & 85 & 420 & 4.8 & 1\\
$\eta$~Car~B & 2 & 3000 & 5.7 & 1\\
WR 22 &   1.9 & 1785 & 1.9 & 2,3 \\ 
WR 24 &   3.4 &   2160 &  5.0 & 2 \\ 
WR 25A &  2.5 & 2480 & 4.9 & 4 \\ 
WR 25B &  0.25 & 2250 & 0.40 & 4 \\ 
HD 93128 &   0.45 &   3355 &  1.6 &   5 \\ 
HD 93129Aa &   2.0 &   3200 &  6.5 &   5 \\ 
HD 93129Ab &   0.61 &   3200 &  0.52 &   5 \\ 
HD 93129B &   0.45 &   3355 &  1.6 &   5 \\ 
HD 93160 &   0.22 &   2560 &  0.46 &  6 \\ 
HD 93160 &   0.22 &   2560 &  0.46 &  6 \\ 
HD 93161Aa &   0.026 &   2100 &  0.036 &  6 \\ 
HD 93161Ab &   0.018 &   1820 &  0.019 &  6 \\ 
HD 93161B &   0.06 &   2460 &  0.12 &  6 \\ 
HD 93205A &   0.25 &   3080 &  0.75 & 6 \\ 
HD 93205B &   0.026 &   2100 &  0.036 & 6 \\ 
HD 93206A &   0.19 &   1510 &  0.14 & 6 \\ 
HD 93206B &   0.05 &   1820 &  0.053 & 6 \\ 
HD 93250 &   0.056 &   3160 &  0.18 &  7,8 \\ 
HD 93403A &   0.34 &   2760 &   0.82 & 6 \\ 
HD 93403B &   0.04 &   2320 &   0.068 & 6 \\ 
\hline\\[-3mm]
Total     &        &        &    35.6 & \\
\hline
\end{tabular}
\tablefoot{Refs.: 1:\citet{Grohea12}, 2:\citet{Hamea06},3:\citet{LC+22}, 4: \citet{Pradhea21}, 5:\citet{Rainea22}, 6: \citet{Smith06}, 7: \citet{Prinjea90}, 8:\citet{NO18}}
\end{center}
\end{table}
\subsection{Comparison of X-ray luminosity to wind power}

X-ray emission arises in the shock-heated intermediate-density gas and due to mixing between the hotter, shocked ejecta and the denser and colder gas phases. The X-ray luminosity can increase sharply after a strong energy increase; up to a factor of several tens is seen after a supernova in the superbubble, decaying over about $10^5\,$yr, as shown by \citet{Krausea14a}. On average, the simulations predict that about $2-3\times10^{-4}$ of the kinetic power injected by massive stars will be emitted in the 0.2 -- 12~keV band, and a fraction of about $10^{-6}$
in the 2 -- 4.5~keV band.

The current kinetic power output by massive stars in the CNC was estimated by \citet{Smith06} to be $L_\mathrm{kin}=3.4\times10^{38}\,$erg$\,$s$^{-1}$. We collect partly updated estimates from the literature in Table~\ref{stars_w} for the stars we detect in X-rays, which, by comparison to \citet{Smith06}, probably contribute the bulk of the kinetic power. Some values have increased in more recent analyses. For our sample only, we already find a combined wind power of  
$L_\mathrm{kin}=3.6\times10^{38}\,$erg$\,$s$^{-1}$.

We derive a thermal X-ray luminosity of the entire CNC of $1.31\times10^{35}\,$erg$\,$s$^{-1}$ (0.2 -- 12~keV), or about 
$\lesssim 3\times10^{-4}$ of the wind power. This is for our adopted assumption for the foreground absorption (without this, the values would be about 40\% higher). This compares very well to the predictions of \citet{Krausea14a}. In the 2 -- 4.5~keV band, we measure a thermal X-ray luminosity of $2.35\times10^{33}$~erg$\,$s$^{-1}$, which corresponds to a fraction of $\lesssim 7\times10^{-6}$ of the wind power. This is again in reasonable agreement with the prediction from the 3D superbubble simulations, and indicates that the thermal structure in the emerging Carina superbubble is similar to the one in the simulations.

A comparison between Table~\ref{stars_w} and Fig.~\ref{stars_image}
confirms that the strongest stellar winds are from the $\eta$~Car binary and the surrounding stars, and that  the gas dynamics observed in X-rays in the CNC is thus indeed driven by this region.

%-----------------------------------------------------------------
\section{Summary}

We have analysed the diffuse X-ray emission in the CNC observed in four all-sky surveys with eROSITA (eRASS:4). The complete coverage of the CNC and its surroundings with eROSITA has allowed us to constrain the local X-ray background emission in a more improved way than in the studies before and to obtain reliable fit results of the X-ray spectra. We created images in X-rays, which were compared to multi-wavelength images, in particular with HI data to identify cavities in the cold ISM. These cavities have most likely been created by the winds of the massive stars and are filled with hot X-ray emitting, low-density plasma. In all regions the dominating emission is modelled best as emission from thermal plasma.
Near the massive stars, in particular, $\eta$ Car, additional non-thermal emission component is detected in the X-ray spectrum.

The most prominent feature in the CNC in X-rays is the V-shaped X-ray hook, which is anti-correlated with the bright triangular shape of photoionised gas seen in the optical.
As the spectral analysis of the eROSITA data have shown, in the X-ray bright Hook region H1 located south of the massive stars, the emission of the hot plasma component is stronger than in other regions.  In this part of the nebula, the ionisation timescale is also high, suggesting high density, which is also consistent with the high brightness. 
The X-ray hook  is most likely the edge of a superbubble formed by the shocks of the stellar winds of massive stars, mainly by $\eta$ Car. The fainter, more diffuse emission in the outer regions is caused by the hot gas which fills the superbubbles around the massive stars in the CNC. 
The pressure inside the superbubbles, as derived from the spectral fit parameters, is one to two orders of magnitude higher than in the ISM. We suggest that the Outer regions result from mixing of the hot gas driven by 
an overpressure in the Central regions through the cold filamentary gas in the Hook region, with the ablated cold gas.

The spectrum in the centre of the CNC is harder than in the outer regions, indicating that there is either an additional hot thermal emission as found by \citet{2011ApJS..194...15T} or  a non-thermal component. The latter is favoured in the fit of the eRASS:4 spectrum in the region around $\eta$ Car and can be explained as synchrotron emission from electrons accelerated in enhanced magnetic fields inside the CNC. From the photon index and the flux obtained from X-ray spectrum we derived the energy distribution and the pressure of the electron gas and a magnetic field strength of $B \approx 60 \mu$G assuming equipartition. The theoretical size of the termination shock around $\eta$ Car is very much consistent with the distance of the X-ray hook from $\eta$ Car. 
The luminosity of the diffuse X-ray emission in the entire Carina nebula is about $10^{35}\,$erg$\,$s$^{-1}$ (0.2 -- 12~keV), which is consistent with the luminosity of the diffuse emission obtained with \chandra\ \citep{2011ApJS..194...16T} and expectations from superbubble simulations \citep{Krausea14a}. 

We have also studied the X-ray emission from the massive stars and binaries in the Carina Nebula, in particular the LBV binary $\eta$ Car and the Wolf-Rayet stars WR~22, WR~24, and WR~25, and bright O stars. Their X-ray spectra are well modelled with one or two components of thermal plasma emission, with relatively high temperatures ($kT > 0.6$~keV). The data of the four surveys show variation of the X-ray flux with orbital phase for $\eta$~Car, WR~22 and WR~25. $\eta$~Car was observed shortly before periastron in eRASS1, with brighter emission above 1~keV. This harder variable emission can be explained as emission from the wind-wind-collision zone in the binary.

\begin{acknowledgements}
In fond memory of Dr.\ Leisa Townsley.
This work is based on data from eROSITA, the soft X-ray instrument aboard SRG, a joint Russian-German science mission supported by the Russian Space Agency (Roskosmos), in the interests of the Russian Academy of Sciences represented by its Space Research Institute (IKI), and the Deutsches Zentrum f\"ur Luft- und Raumfahrt (DLR). The SRG spacecraft was built by Lavochkin Association (NPOL) and its subcontractors, and is operated by NPOL with support from the Max Planck Institute for Extraterrestrial Physics (MPE).

The development and construction of the eROSITA X-ray instrument was led by MPE, with contributions from the Dr. Karl Remeis Observatory Bamberg \& ECAP (FAU Erlangen-N\"urnberg), the University of Hamburg Observatory, the Leibniz Institute for Astrophysics Potsdam (AIP), and the Institute for Astronomy and Astrophysics of the University of T\"ubingenn, with the support of DLR and the Max Planck Society. The Argelander Institute for Astronomy of the University of Bonn and the Ludwig Maximilians Universit\"at Munich also participated in the science preparation for eROSITA.
The eROSITA data shown here were processed using the eSASS/NRTA software system developed by the German eROSITA consortium.
The Digitized Sky Surveys were produced at the Space Telescope Science Institute under U.S. Government grant NAG W-2166. The images of these surveys are based on photographic data obtained using the Oschin Schmidt Telescope on Palomar Mountain and the UK Schmidt Telescope. The plates were processed into the present compressed digital form with the permission of these institutions. 
M.S. acknowledges support from the Deutsche Forschungsgemeinschaft through the grants
SA 2131/13-1, SA 2131/14-1, and SA 2131/15-1.
J.R. acknowledges support from the DLR under grant 50QR2105.
\end{acknowledgements}

\bibliographystyle{aa}
\bibliography{refs.bib}

\begin{appendix}

\section{eRASS catalogue entries}
\label{app_stars}

\begin{table}[h]
\caption{\label{stars_id}eRASS1-data release 1 (DR1) main catalogue IDs of the analysed point sources}
\begin{center}
\begin{tabular}{lr}\hline\hline\\[-3.mm]
main source(s) & 1eRASS \\\hline
WR 22 (HD 92740)          &  J104117.5-594036  \\
Cl* Collinder 228 TM 103  &  J104246.5-601209  \\
WR 24 (HD 93131)          &  J104352.3-600704  \\
HD 93129 + HD 93128       &  J104356.9-593252  \\
HD 93160 + HD 93161       &  J104408.3-593432  \\
WR 25 (HD 93162)          &  J104410.2-594310  \\
HD 93206 (QZ Car)         &  J104422.8-595936  \\
HD 93205 (V560 Car)       &  J104433.7-594412  \\
HD 93250                  &  J104445.0-593355  \\
$\eta$ Car (HD 93308)     &  J104503.4-594103  \\
HD 93403                  &  J104544.0-592427  \\\hline
\end{tabular}
\end{center}
\end{table}

\end{appendix}

\end{document}